\documentclass[prb,superscriptaddress,longbibliography,twocolumn]{revtex4-1}
\pdfoutput=1
\usepackage{epsfig}
\usepackage{geometry}
\usepackage{verbatim}
\usepackage{amsmath}
\usepackage{amssymb}
\usepackage{graphicx,picture,calc}
\usepackage{hyperref}
\usepackage{braket}
\usepackage{bm}
\usepackage{epstopdf}
\usepackage[normalem]{ulem}
\usepackage[usenames,dvipsnames]{color}
\usepackage[multidot]{grffile}
\usepackage{overpic}
\usepackage{dsfont}
\newcommand{\be}{\begin{equation}}
\newcommand{\ee}{\end{equation}}
\newcommand{\bk}{{{\bf{k}}}}

\newcommand{\bQ}{{{\bf{Q}}}}

\newcommand{\bea}{\begin{eqnarray}}
\newcommand{\eea}{\end{eqnarray}}
\newcommand{\beal}{\begin{align}}
\newcommand{\eeal}{\end{align}}
\renewcommand{\i}{{\i}}

\renewcommand{\i}{{\mathrm{i}}}
\newcommand{\e}{{\mathrm{e}}}

\newcommand{\dd}{\mathrm{\mathbf{d}}}
\newcommand{\Tr}{\mathrm{Tr}}
\newcommand{\tr}{\mathrm{tr}}

\graphicspath{{figures/compressed/}}

\begin{document}

\title{Molecular dipoles in designer honeycomb lattices}

\author{Nazim Boudjada}
\affiliation{Department of Physics, University of Toronto, Toronto, Ontario M5S1A7, Canada.}
\author{Finn Lasse Buessen}
\affiliation{Department of Physics, University of Toronto, Toronto, Ontario M5S1A7, Canada.}
\author{Arun Paramekanti}
\email{arunp@physics.utoronto.ca}
\affiliation{Department of Physics, University of Toronto, Toronto, Ontario M5S1A7, Canada.}

\date{\today}

\begin{abstract}
Recent advances in ultracold atoms in optical lattices and developments in surface science
have allowed for the creation of artificial lattices as well as the control of many-body interactions. 
Such systems provide new settings to investigate interaction-driven instabilities and non-trivial topology. In this paper, we explore the interplay between molecular electric dipoles on a 
two-dimensional triangular lattice with fermions hopping on the dual decorated honeycomb lattice which hosts Dirac and flat band states. 
We show that short-range dipole-dipole interaction can lead to ordering into various stripe and vortex crystal ground states. We study these ordered states and
their thermal transitions 
as a function 
of the interaction range using simulated annealing and Monte Carlo methods. For the special case of zero wave vector ferrodipolar order, incorporating dipole-electron interactions and integrating out the electrons leads to a six-state clock model for the dipole ordering. Finally, we discuss the impact of the various dipole orders on the 
electronic band structure and the local tunneling density of states. 
Our work may be relevant to studies of ``molecular graphene'' --- CO molecules arranged on the Cu(111) surface --- which have been explored using scanning 
tunneling spectroscopy, as well as ultracold molecule-fermion mixtures in optical lattices.
\end{abstract}

\maketitle

\section{Introduction}
Magnetic interactions in materials are usually modeled by Heisenberg-like $J_{ij}\vec{S}_i\cdot\vec{S}_j$ Hamiltonians with $J_{ij}$ couplings restricted to a few nearest-neighbors, which can be calculated perturbatively or using {\it ab initio} methods, or deduced from fits to experiments. However, the extension to longer-range
and anisotropic
spin-exchange Hamiltonians due to entangling of spin and spatial degrees of freedom is often necessary in many different contexts\cite{mckagome,bosegas,frgdipoles,hinokihara2020phase,dipoleskagome}. In frustrated spin-ice systems such as the rare-earth
pyrochlore oxides \cite{pyrochlore}, the large magnetic dipole moments lead to an appreciable energy scale $\sim 1$\,K for the magnetic dipole-dipole interaction,
which has to be taken into account to capture their Curie-Weiss temperatures. In ultracold optical lattices, Rydberg atoms can be used as quantum simulators for many-body physics by mapping the spin degree of freedom to the population of different excited Rydberg states \cite{rydberg,qcprydberg,rydbergkagome,rydbergkagome2} and result in both a density-density interaction term with $V_{ij}\sim1/|\vec{R}_{ij}|^6$ and XY spin models with an energy scale $J_{ij}\sim1/|\vec{R}_{ij}|^3$. Engineered Hamiltonians with polar molecules have also been predicted to lead to long-range order in two dimensions (2D) \cite{LRO} and to harbor quantum spin liquids in the triangular and kagome lattices\cite{quantumdipoles,frgdipoles}. 

The directional
dipolar interaction leads to frustration effects \cite{magthin}, similar to the bond-directional Kitaev couplings which support 
unconventional spin liquids or large-scale spin textures \cite{jackeli2009,reviewkitaev1,reviewkitaev2,kitaevqsl,kitaevmagnon,liern,batista2019}.
In recent years, such dipolar models have also become of interest in systems supporting electric dipole moments, which include
Mott insulators of organic molecules \cite{Drichko2018}, honeycomb Kitaev magnets \cite{Loidl2018},
``molecular graphene'' which consists of carbon monoxide (CO) molecules arranged on a Cu(111) surface to form a triangular lattice \cite{Manoharan,Polini},
and possibly the inversion broken surface of nearly ferroelectric materials or of dichalcogenides such as {1T-TaS$_2$} in its charge-ordered phase.
By tuning the distance between the dipoles, one can control the strength of the dipole-dipole interaction, which can also be modified by screening from the underlying substrate. Moreover, for molecular graphene and {1T-TaS$_2$}, we can have electrons which live on a dual decorated honeycomb lattice which can 
simulate some of the physics of graphene and flat band systems.
Experiments have shown that Dirac fermions emerge in molecular graphene, as well as in other artificial lattices realized using ultracold atoms \cite{Tarruell2012} or semiconductor devices \cite{agsemiconductor}.
These systems offer a versatile platform to explore different phases of matter which are usually inaccessible in conventional graphene. For example, spin-orbit coupling can be rather large, many-body effects can be tuned, and pseudo gauge fields can be easily generated using strain. 
The dual honeycomb lattice for the electrons may also host additional intermediate sites similar to the Lieb lattice
\cite{honeycombflatband,liebsquare}; in this case, flat electronic bands stemming from destructive interference between the wavefunctions on different lattice sites emerge, 
and are robust against small perturbations and the addition of further neighbor hoppings. 
Such flat bands can lead to non-trivial topology \cite{liebtopology,cdwhoneycomb,honeycombwires} and symmetry broken states 
such as $p+\i\tau p$ Fulde-Ferrell-Larkin-Ovchinnikov (FFLO) states \cite{superconductivityAG}, 
$d+\i d$ superconductivity and charge density order \cite{honeycombwires,cdwhoneycomb}, or Wigner crystals \cite{wignercrystal}. 
When the dipolar molecules are more densely packed, the enhanced dipole-dipole interaction can lead to new broken symmetry states with 
a greatly enlarged unit cell. Similarly to the physics of moir\'e superlattices in twisted bilayer graphene, the flat electronic bands may give rise to a strong-coupling picture \cite{tbg}.

This paper is organized as follows. In Sec. \ref{sec:DDI} we start by considering spatially extended interactions between molecular dipoles on a triangular lattice. Using 
variational calculations and  simulated annealing methods, we uncover the various types of ordered states which emerge as we include dipolar interactions
that are cut off with a range parameter which can be viewed as a crude way to mimic screening effects from the substrate. 
Although previous studies using Ewald summation techniques have shown that the ground state in the thermodynamic limit with infinite range dipolar interactions has in-plane ferrodipolar order \cite{triangularmc}, we find that imposing a finite range cutoff leads to a rich set of stripe ordered states or proximate multi-$Q$ vortex crystals.
We also study the finite temperature phase transitions upon heating such stripe or vortex crystal phases using Monte Carlo simulations.
 In Sec. \ref{sec:DF}, we consider the Dirac fermions which form the dual honeycomb lattice and model the hopping problem of a decorated lattice with an arbitrary number of additional sites along the bonds. In the case of the ferro aligned dipole moments, we show that order by disorder will pin the dipole moments either along or perpendicularly to the triangular lattice direction. Then, we include the electron-dipole coupling, we perform tight-binding calculations to investigate the impact of the various dipole orders on the electronic band structure, and we discuss implications for scanning tunneling spectroscopy (STS) experiments on the
 broken symmetry states.

\section{Dipole-dipole interactions}
\label{sec:DDI}

\begin{figure}[t]
	\centering
	\includegraphics[width=0.47\linewidth]{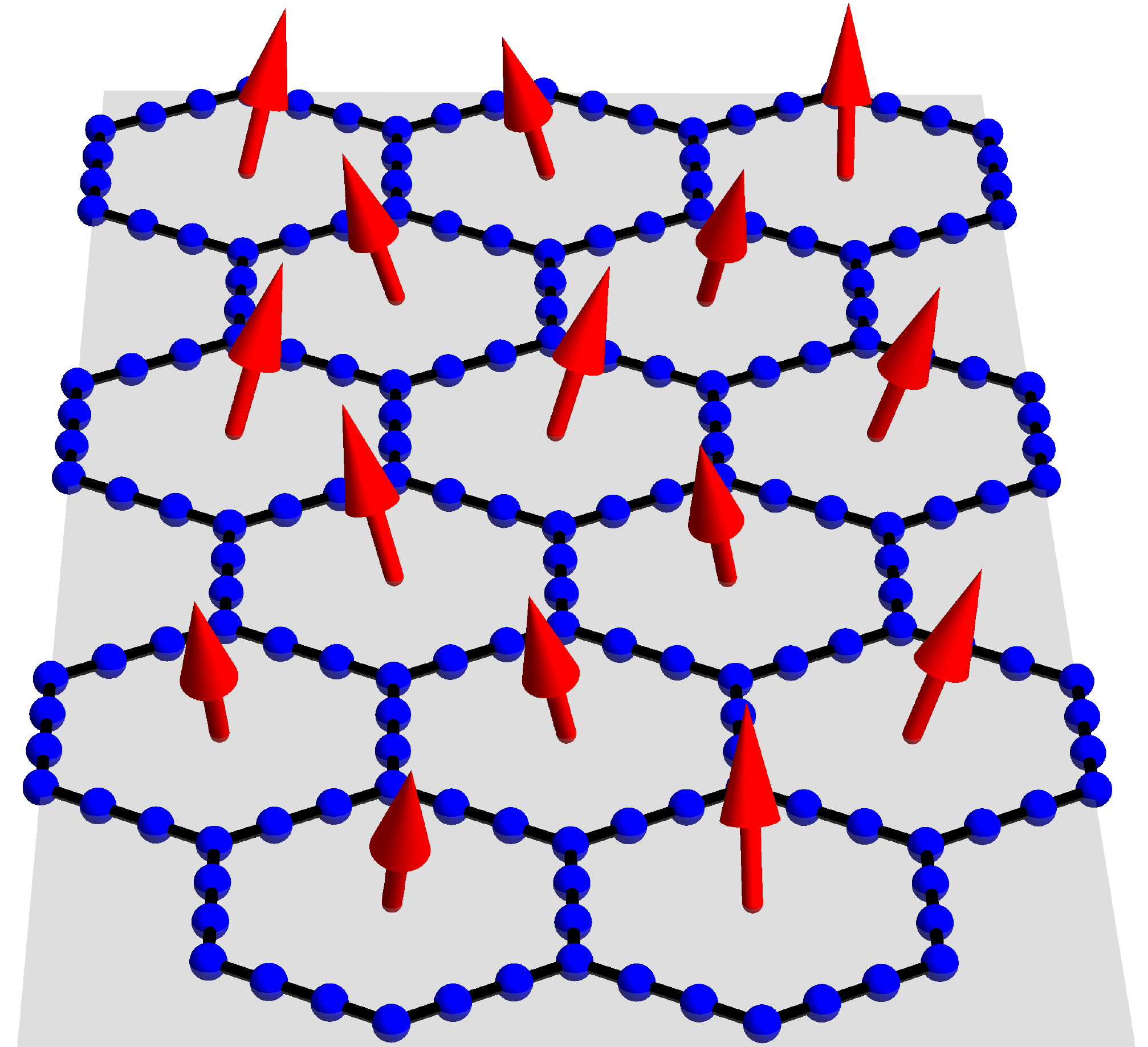}
	\includegraphics[width=0.51\linewidth]{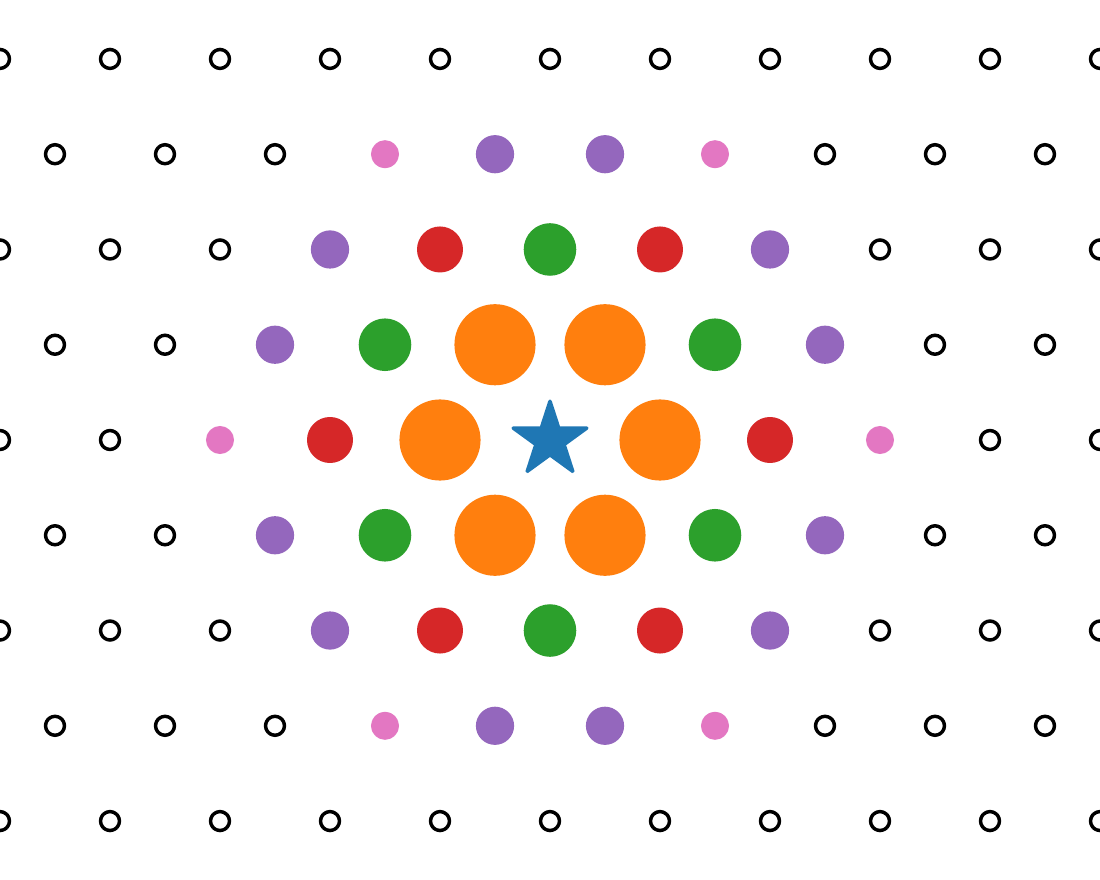}
	\caption{Left: Electric dipoles on hexagonal plaquettes of a decorated honeycomb lattice. Right: Triangular lattice with a central site (blue star) and its $n$th nearest-neighbors: the 6 first nearest-neighbors are shown in orange, the 6 second nearest-neighbors are shown in green, the 6 third nearest-neighbors are shown in red, the 12 fourth nearest-neighbors are shown in violet, the 6 fifth nearest-neighbors are shown in pink, and the rest are shown as open circles. }
	\label{fig:NN}
\end{figure}

We start by considering the classical long-range electric dipole-dipole interaction,
\begin{equation}
H_d\!=\!\sum_{i < j}\frac{1}{4\pi\epsilon_0|\vec{R}_{ij}|^3}\left[\dd_i \!\cdot\!\dd_j \!-\! 3\; (\dd_i \!\cdot\! \hat{R}_{ij}) \; (\dd_j \!\cdot\! \hat{R}_{ij}) \right],
\label{eqn:dipH}
\end{equation}
where $\vec{R}_{ij}$ connects dipoles $\dd_i$ and $\dd_j$ which live on a triangular lattice. We fix the triangular lattice constant $a \!\equiv\! 1$, and
measure all couplings in terms of the nearest-neighbor dipole interaction strength $J_1 \equiv |\dd|^2/(4\pi\epsilon_0 a^3)$ where $|\dd|$ is the dipole strength. 
The decay of the dipolar interaction
with distance implies that the $n$th neighbor coupling $J_n = J_1/|\vec{\delta}_n|^3$, where $|\vec{\delta}_n|$ is the distance to the $n$th neighbor, in terms of the
lattice constant. For instance, $J_2/J_1\! \equiv \! 3^{-3/2}\! \approx \! 0.19$ and $J_3/J_1 \! \equiv \! 2^{-3} \!=\! 0.125$.
Figure ~\ref{fig:NN} shows the full set of nearest-neighbors up to $\vec{\delta}_5$. For instance,
\begin{equation}
\vec{\delta}_1=\begin{cases} \pm\left(1,0\right)\\ \pm\left(\frac{\sqrt{3}}{2},\frac{1}{2}\right)\\ \pm\left(-\frac{\sqrt{3}}{2},\frac{1}{2}\right) \end{cases}\hspace{-0.5cm},~~~~
\vec{\delta}_2=\begin{cases}\pm\sqrt{3}\left(0,1\right)\\\pm\sqrt{3}\left(\frac{1}{2},\frac{\sqrt{3}}{2}\right)\\\pm\sqrt{3}\left(-\frac{1}{2},\frac{\sqrt{3}}{2}\right)\end{cases}.
\end{equation}
Below, we will explore the impact of such dipole interactions $J_n$ when we impose a cutoff at a neighbor range $n_{\max}$ (which we will vary). For $n_{\max}=5$, a total of 36 dipoles will interact with each given dipole and system size effects can become dominant for the system sizes we consider in this paper.
Small $n_{\max}$ values crudely mimic the impact of screening, while $n_{\max} \to \infty$ incorporates the full long-range dipole interaction.

\subsection{Zero temperature orders}

We study the model Hamiltonian with different cutoffs $n_{\max}$ using Luttinger-Tisza, classical Monte Carlo, or 
variational calculations. The Luttinger-Tisza method provides quick insights : We Fourier-transform the Hamiltonian to momentum space to obtain
\begin{eqnarray}
H_d&=&\sum_{n,\vec{\delta}_n,\bk} J_n\left[\dd^*_{\bk}\cdot\dd_{\bk}-3 (\dd^*_{\bk}\cdot \hat{\delta}_{n})(\dd_{\bk}\cdot \hat{\delta}_{n})\right]\cos(\bk\cdot\vec{\delta}_{n})\nonumber\\
&=&\sum_{n,\vec{\delta}_{n},\bk} \frac{\cos(\bk\cdot\vec{\delta}_{n})}{|\vec{\delta}_{n}|^3} ~~ {\dd^{\alpha *}_{\bk}} \dd^\beta_{\bk} \!\!\!
\underbrace{(\delta_{\alpha,\beta}-3 \hat{\delta}_n^\alpha \hat{\delta}_{n}^\beta),}_{\begin{bmatrix}
	1 & 0 & 0\\
	0 & 1 & 0\\
	0 & 0 & 1
	\end{bmatrix} - 3
	\begin{bmatrix}
	* & * & 0\\
	* & * & 0\\
	0 & 0 & 0
	\end{bmatrix}}
\end{eqnarray}
The structure of this matrix shows that the in-plane ($xy$) and out-of-plane ($z$) dipole moments are decoupled. We can thus consider the normal component of the dipole (perpendicular to the triangular layer) to be fixed, say by an external electric field, while the in-plane component can independently order because of the dipolar interactions. 
This in-plane order is what we find in our simulations discussed below.
The minimal eigenvalues of this Luttinger-Tisza matrix over all $\bk$ yield the dipole ordering wave vector.
However, while the Luttinger-Tisza method gives reasonable predictions for the energy and the wave vector which minimizes the Hamiltonian, it does not always satisfy the hard spin constraint. We therefore focus below on discussing the Monte Carlo and variational calculations.

We consider the classical model of Eq. \eqref{eqn:dipH} by mapping the dipole moments $\dd$ to classical $O(3)$ vectors and we minimize $H_d$ at $T=0$ using simulated annealing. We start with only the nearest-neighbors (i.e., setting $n_{\max}=1$) and then successively incorporate further neighbors to study how it impacts the dipole configurations. We also tune the couplings away from the purely dipolar constants to explore nearby phases. In each case, we calculate the ordering wave vector from the peak of the static structure factor which is the Fourier transform of the spin-spin correlation function:
\begin{equation}
S(\bk)=\frac{1}{N_d}\sum_{i,j}\e^{\i\bk\cdot(\vec{R}_i-\vec{R}_j)}\langle \dd_i\cdot\dd_j\rangle,
\end{equation}
where $N_d$ is the total number of dipoles. Our results indicate two families of orders: stripe orders and vortex crystals. We note that although the following results seem to indicate that stripe orders are associated with odd $n_{\max}$ and vortex orders with even $n_{\max}$, in reality this observation is caused by geometric frustration effects induced by the outermost ring of neighbors considered. For example, when the outermost neighbors lie on the $x$ axis 
(and corresponding $\pi/3$ rotated sites), such as the orange, red, and pink neighbors in Fig. \ref{fig:NN}), we find a $\bQ$ on the $\Gamma\rightarrow\mathrm{M}$ direction in the Brillouin zone (BZ).

\subsubsection{Stripe orders}

\begin{figure}[t]
	\centering
	\hspace{-0.5cm}\includegraphics[width=\linewidth]{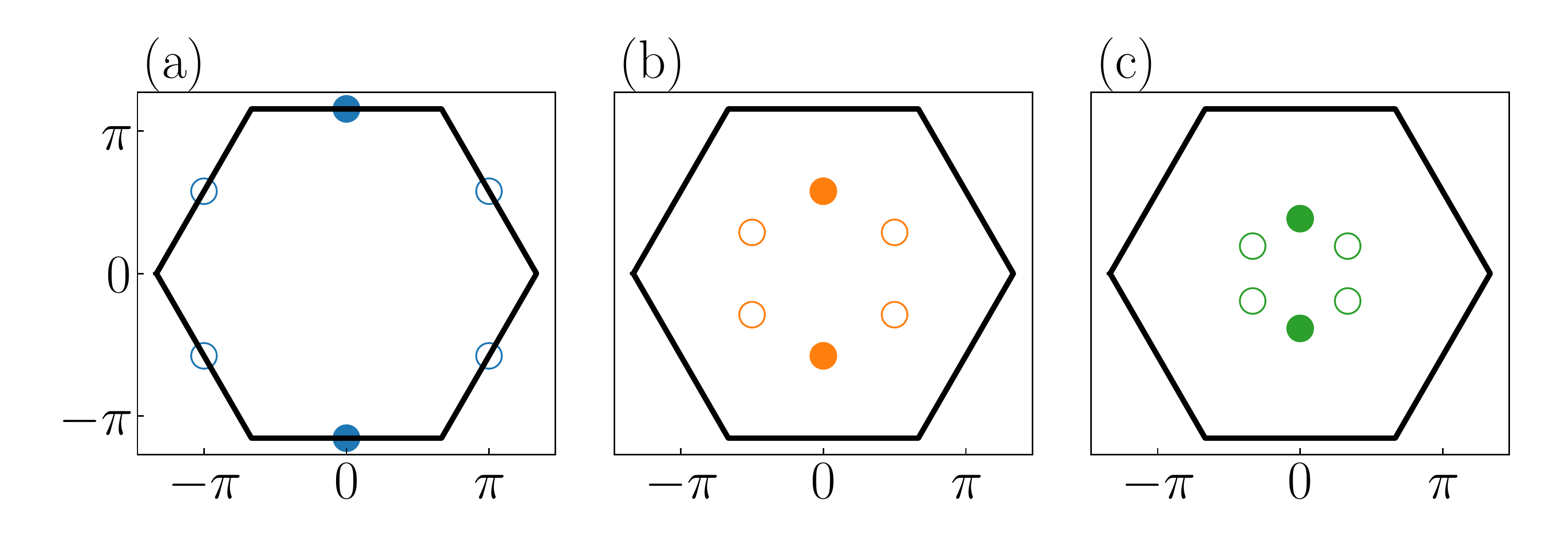}
	\hspace{0.75cm}\includegraphics[width=0.26\linewidth]{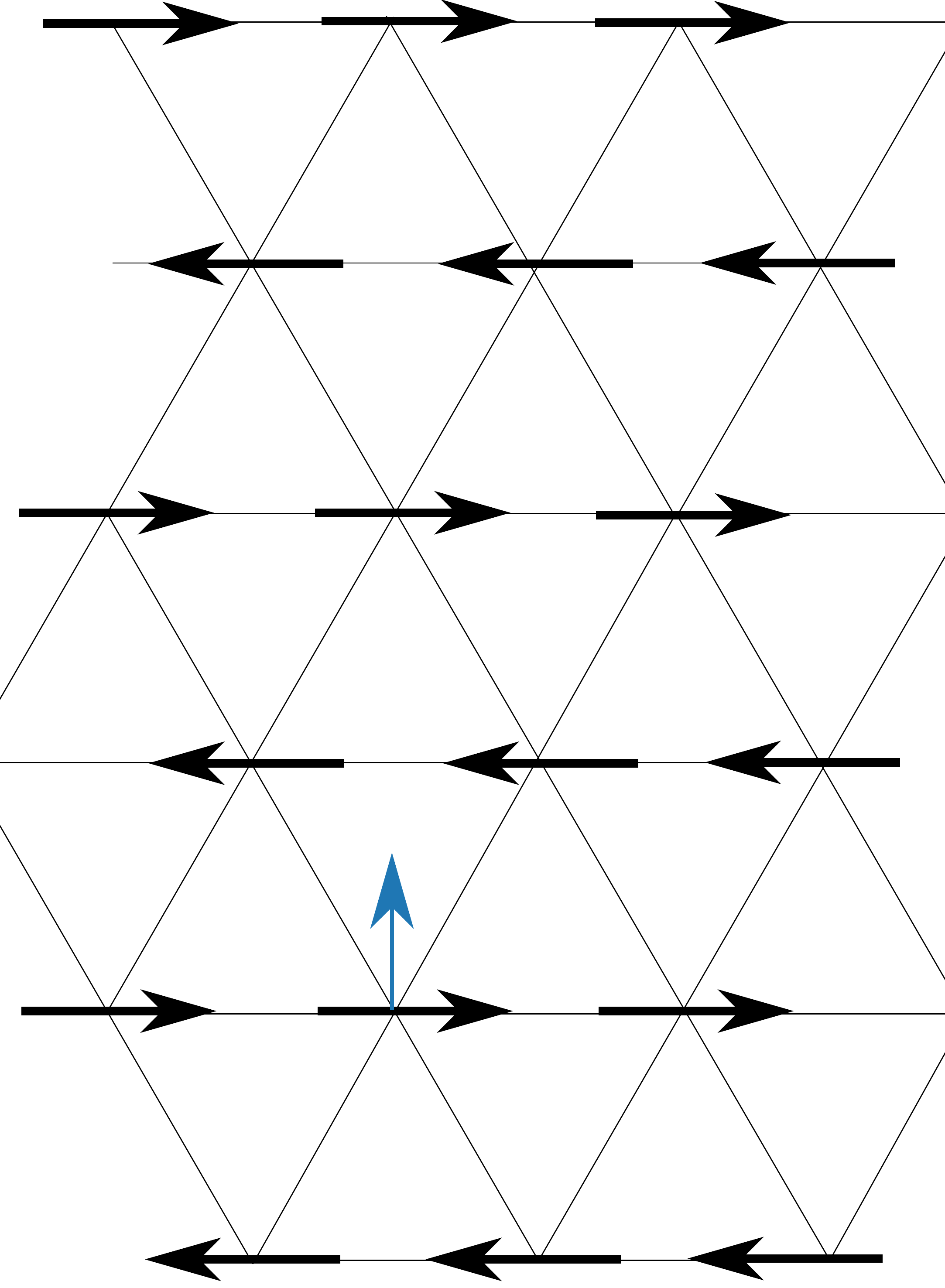}\hspace{0.2cm}\vrule\hspace{0.2cm}
	\includegraphics[width=0.26\linewidth]{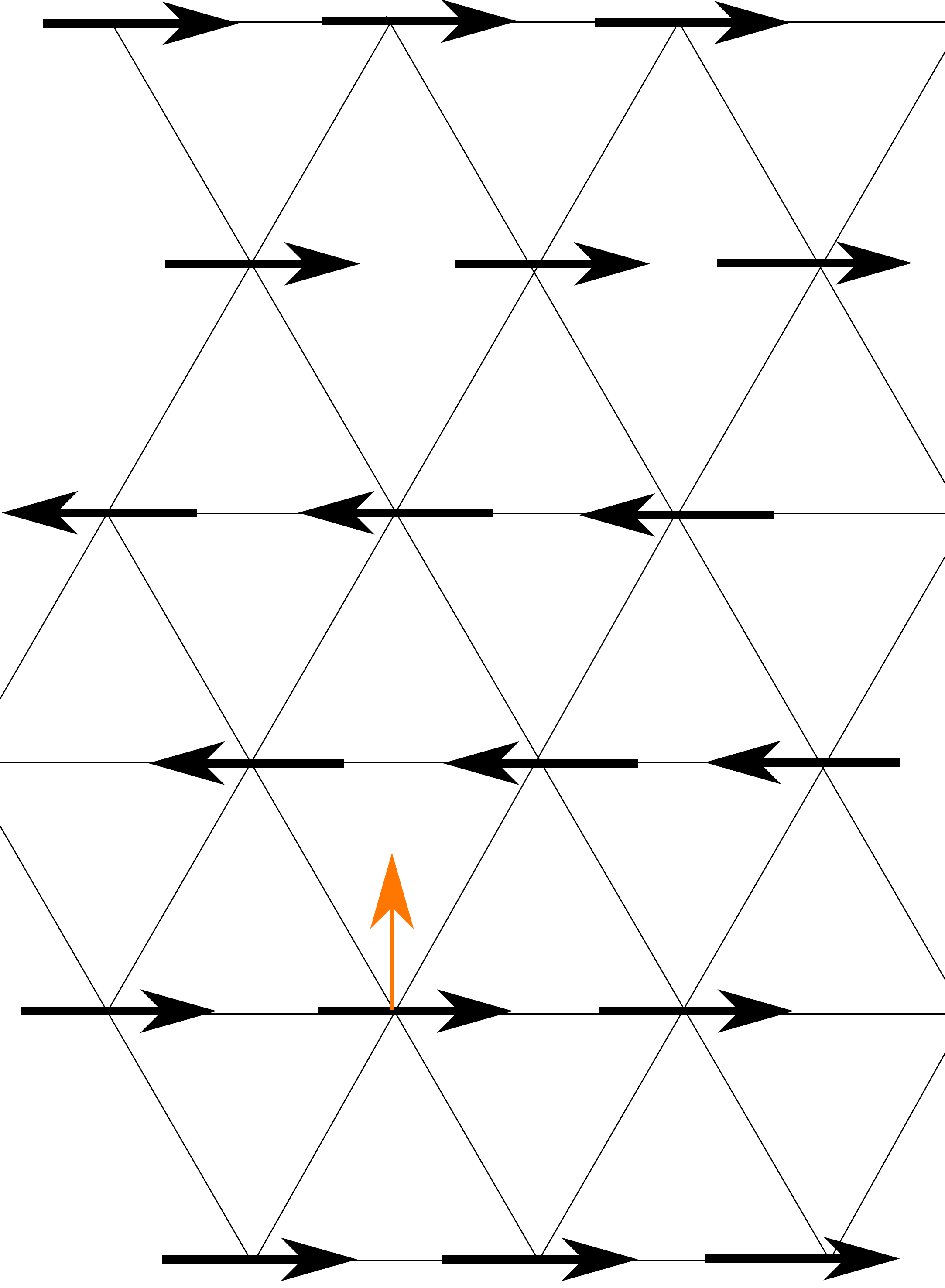}\hspace{0.2cm}\vrule\hspace{0.2cm}
	\includegraphics[width=0.26\linewidth]{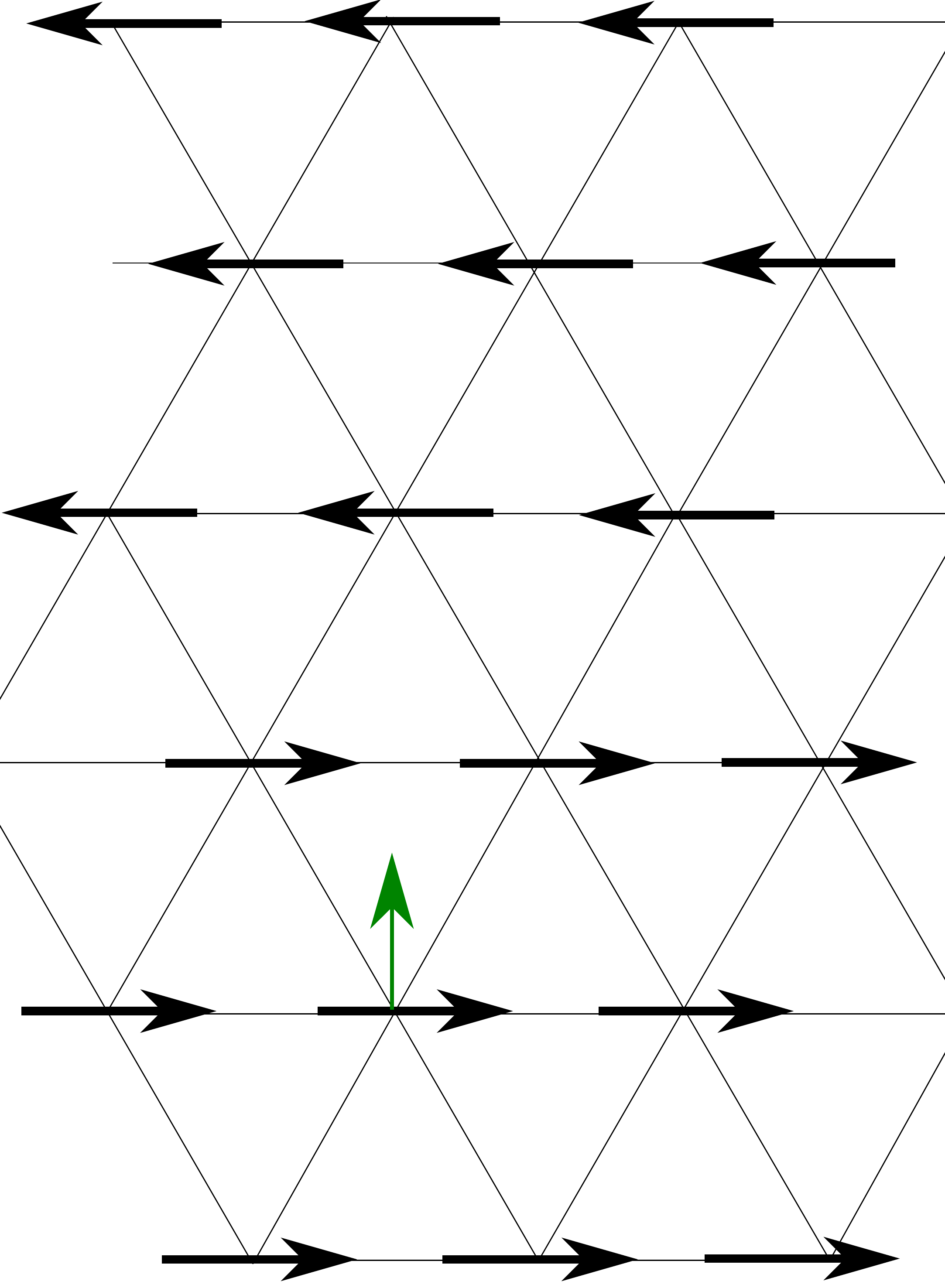}
	\caption{Top: Peaks of the static structure factor for stripe orders obtained for (a) $n_{\max}=1$ with $\bQ$ at the M points, (b) $n_{\max}=3$ with $\bQ$ at M/2, and (c) $n_{\max}=5$ with $\bQ$ at M/3. Bottom: Real-space configurations corresponding to the filled circles (i.e., $\bQ\propto\hat{y}$), although all $C_6$ rotations of these are energetically equivalent. $n_{\max}=1$ (a), $n_{\max}=3$ (b) and $n_{\max}=5$ (c) corresponding to the filled circles of the top panels.}
	\label{fig:sfdip}
\end{figure}

We find stripe orders with different ordering wave vectors $\bk=\bQ$ as a function of $n_{\max}$. In the top panels of Fig.~\ref{fig:sfdip} we show the peaks of $S(\bk)$ in the first Brillouin zone (FBZ), with the Real-space pictures corresponding to the filled $\bQ$ circle in the bottom panels. For example, when only considering the nearest-neighbor interaction, we find $|\bQ|=2\pi/\sqrt{3}$ at the M points of the BZ but the Monte Carlo simulation spontaneously picks one of the six energetically equivalent M points (filled circles) as shown in Fig.~\ref{fig:sfdip}(a). The real-space picture corresponds to a stripe of dipoles oriented along the direction orthogonal to $\bQ$ (i.e., $\dd\propto\hat{z}\times\bQ$). Similarly, for $n_{\max}=3$, we find a stripe order with $|\bQ|=\pi/\sqrt{3}$, which means the wavelength is doubled as seen in Fig.~\ref{fig:sfdip}(b). For the spin configuration $n_{\max}=5$ we find $|\bQ|=\pi/(2\sqrt{3})$ with a tripled wavelength. In general, we see that stripe orders with arbitrarily long wavelength can be stabilized with $|\bQ|\sim2\mathrm{M}/(n_{\max}+1)$, consistent with the true ferro ground state for the full dipolar Hamiltonian. In all cases, the molecules spontaneously break the $C_6$ and a subset of translational symmetries of the lattice.

\subsubsection{Vortex crystals}

\begin{figure}[!t]
	\centering
	\includegraphics[width=\linewidth]{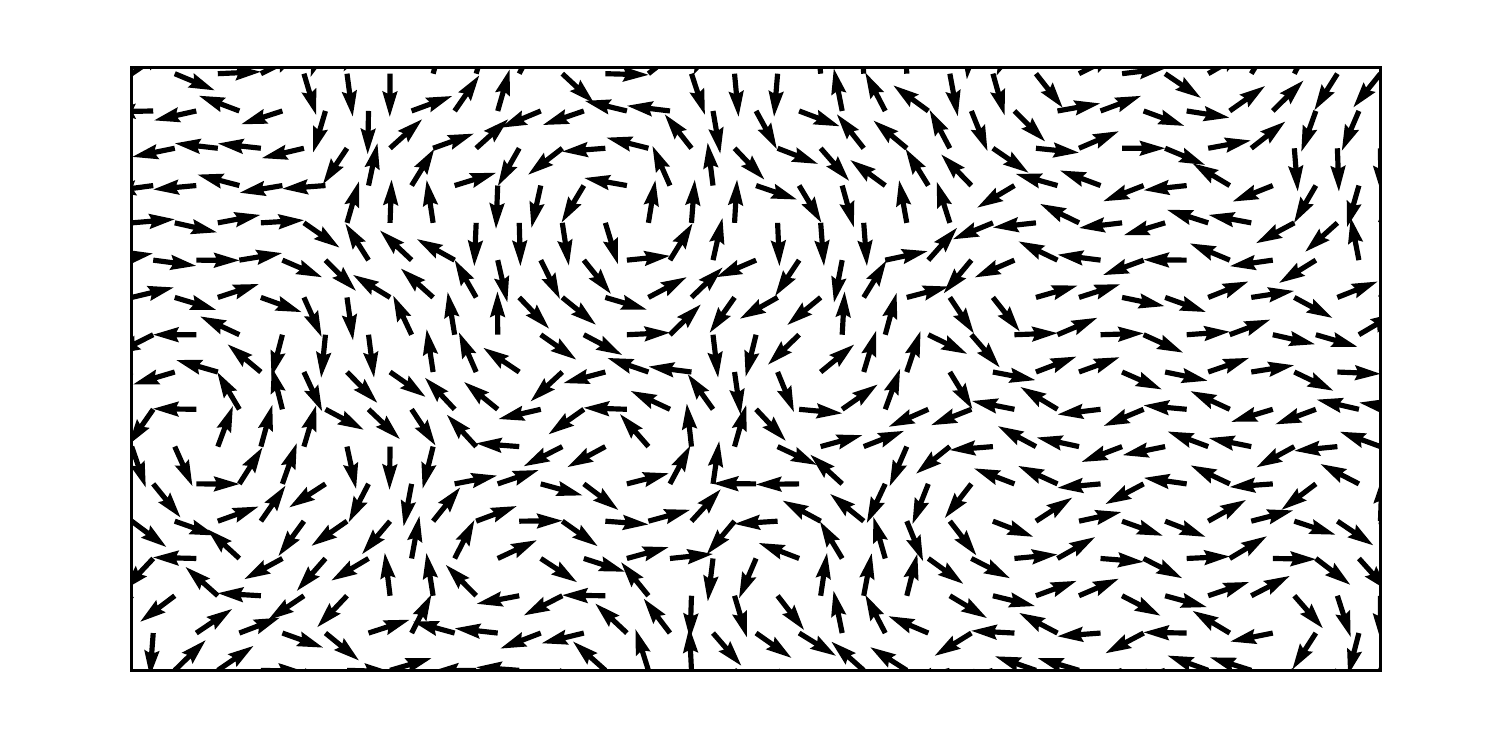}
	\caption{Zero-temperature configuration with $n_{\mathrm{max}}=4$ and dipolar couplings $(J_1,J_2,J_3,J_4)\approx(1,0.19,0.125,0.054)$ showing domains of stripes and non crystallized vortices.}
	\label{fig:dipolar}
\end{figure}

\begin{figure}[t]
	\centering
	\includegraphics[width=\linewidth]{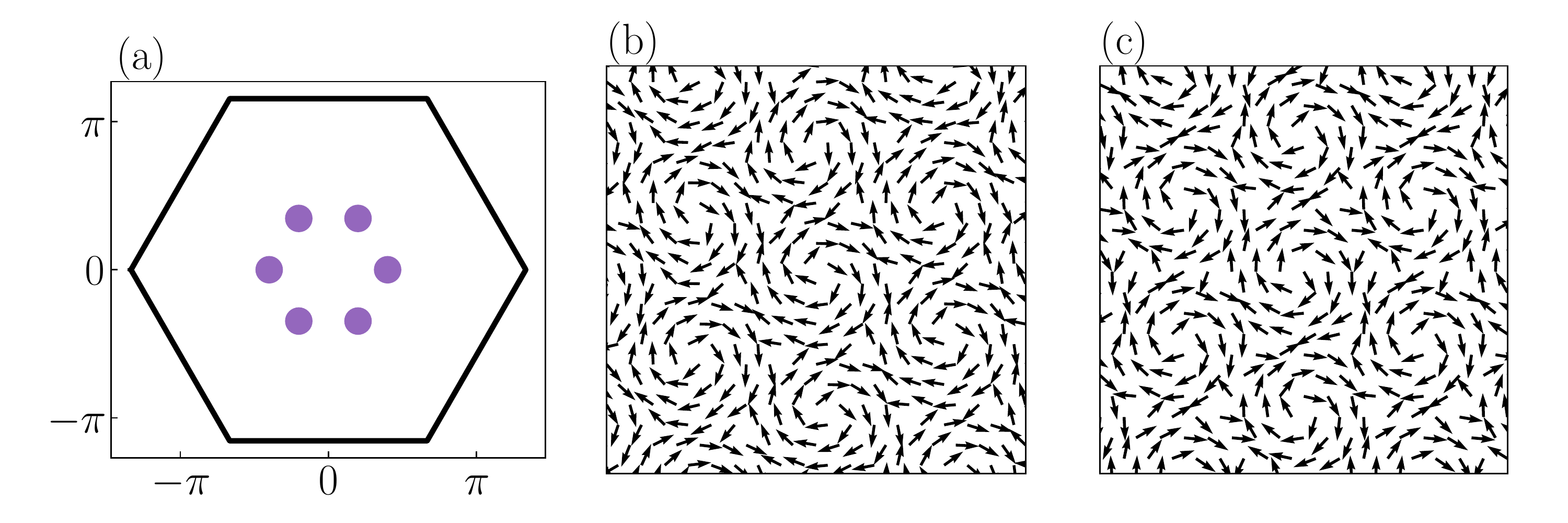}
	\caption{(a) Peaks of the static structure factor at $|\bQ|\approx2\pi/5$ for the vortex order with purely dipolar $J_{1,2,3}$ and $J_4=0.1$. In this multi-$Q$ state, modes with
	$(\bQ_1,\bQ_2,\bQ_3)$ have equal amplitude which leads to a $C_6$ symmetric dipolar configuration. (b) Real-space picture of the vortex/antivortex crystal using classical Monte Carlo yielding $\langle H_d\rangle_{\mathrm{MC}}/N_d=-2.59|J_1|$, and (c) minimum energy configuration obtained from a variational ansatz with ${(|\bQ|,\phi_1,\phi_2,\phi_3)\approx(2\pi/5,4.28,5.34,4.52)}$ yielding $\langle H_d\rangle_{\mathrm{V}}/N_d=-2.53|J_1|$. }
	\label{fig:vortex}
\end{figure}

When $n_{\max}$ is even, we find multi-$Q$ in-plane orders which do not perfectly crystallize as can be seen in Fig.~\ref{fig:dipolar}. They exhibit features which resemble the stripe phase as well as regions marked by the appearance of vortices. It is possible that such structures reflect the close energetic competition between stripe and vortex crystal phases, indicating the presence of a critical point separating them.
We have found that slightly tuning the couplings away from the dipolar constants can stabilize either the single-$Q$ stripe phase or a perfect multi-$Q$ vortex/antivortex crystal in the ground state. As an example, we show the case of $n_{\max}=4$, where the $1/|\vec{\delta}_n|^3$ behavior would predict ${(J_1,J_2,J_3,J_4)\approx(1,0.19,0.125,0.054)}$ (i.e., purely dipolar), but we tune $J_4$ to $0.1$ instead. In Fig.~\ref{fig:vortex}(a), we show the structure factor which peaks simultaneously at six $C_6$-related momenta of magnitude $\bQ\approx2\pi/5$, but now along the $\Gamma\rightarrow\mathrm{K}$ directions. The corresponding real-space configuration is shown in Fig.~\ref{fig:vortex}(b) on a slab of a bigger system of size $60\times60$. 

Since the structure factor peaks equally at all symmetry-related momenta, we propose the following ansatz to describe the dipoles at each site:
\begin{equation}
\dd_i= \frac{1}{\mathcal{N}(\vec{R}_i)} \sum_{\mu=1}^{3} \hat{z}\times \bQ_\mu \cos(\bQ_\mu\cdot\vec{R}_i+\phi_\mu),
\label{eqn:va}
\end{equation}
where $\mathcal{R}_z(2\pi/3)\bQ_{\mu}=\bQ_{\mu+1}$ with $\mathcal{R}_z(\theta)$ being the rotation matrix around the $z$ axis by angle $\theta$, and $\mathcal{N}(\vec{R}_i)$ being a site-dependent normalization constant. We use Eq. \eqref{eqn:va} as a variational ansatz and minimize $H_d$ with respect to the four free parameters to adjust ($|\bQ|,\phi_1,\phi_2,\phi_3$). Our results show very good agreements between the variational minimization and classical Monte Carlo calculations as can be seen from Fig.~\ref{fig:vortex}: In both cases, $|\bQ|\approx2\pi/5$, and the energies are very close: $\langle H_d\rangle_{\mathrm{MC}}/N_d\approx-2.59|J_1|$, while $\langle H_d\rangle_{\mathrm{V}}/N_d\approx-2.53|J_1|$.

\subsection{Thermal phase transitions}
\begin{figure}[!ht]
	\centering
	\includegraphics[width=\linewidth]{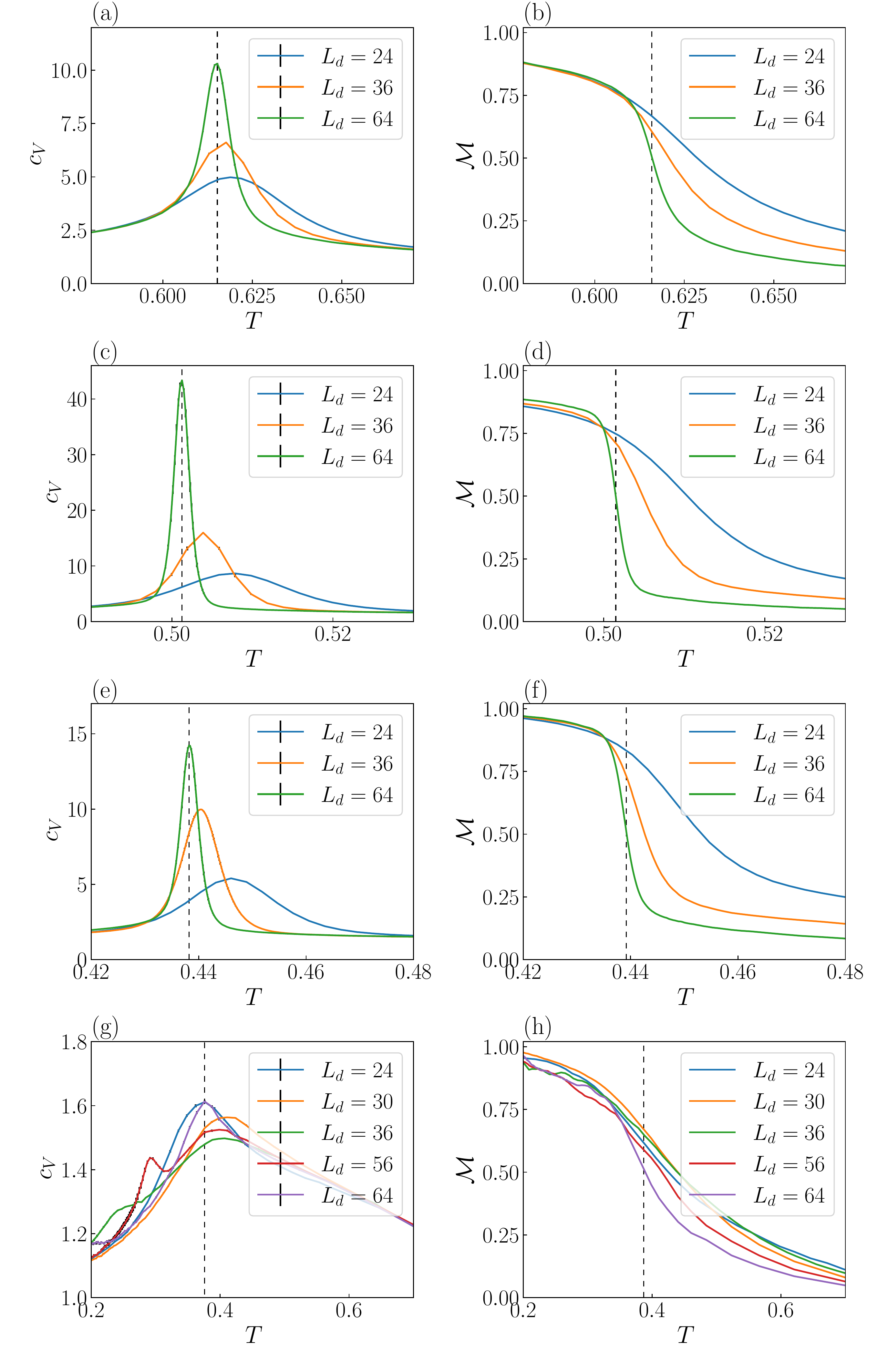}
	\caption{Specific heat $c_V$ (left) and the corresponding order parameter $\mathcal{M}$ (right) as a function of temperature for the stripe-$n^{\mathrm{max}}$ and vortex phases as obtained from classical Monte Carlo simulations using the parallel tempering algorithm with dashed lines showing the transition temperature in the largest system size and black lines indicating the error bars. (a) and (b) $n^{\mathrm{max}}=1$ with ordering wave vector at the M points, (c) and (d) $n^{\mathrm{max}}=3$ with ordering wave vector at the M/2 points and, (e) and (f) $n^{\mathrm{max}}=5$ with ordering wave vector at the M/3 points, and (g) and (h) vortex crystal with ordering wave vector $2\pi/5$. }
	\label{fig:cvop}
\end{figure}

In order to study the thermal phase transitions into the ordered ground states, we use finite temperature Monte Carlo simulations with parallel tempering to efficiently explore the free energy landscape. For the stripe orders, we run $10^6$ thermalization sweeps and $9\times10^6$ measurement sweeps for system sizes $L_d\in[24,36,64]$ and perform replica exchanges every $10$ sweeps with $144$ temperature points logarithmically distributed between $T_{\mathrm{max}}=5$ and $T_{\mathrm{min}}=0.05$. In the left panels of Fig.~\ref{fig:cvop}, we show the heat capacity per spin $c_V(T)=\frac{\langle H_d^2\rangle-\langle H_d\rangle^2}{N_dT^2}$ for the stripe-1, stripe-2, stripe-3, and vortex crystal phases. We find very sharp peaks for the stripe phases which can be extrapolated using finite-size scaling to the $L_d\rightarrow\infty$ limit and extract $T_{c,s1}\approx0.612$, $T_{c,s2}\approx0.497$, and $T_{c,s3}\approx0.433$. For the vortex crystal, we use $2\times10^6$ thermalization sweeps and $20\times10^6$ measurement sweeps. In this case,
we find broad heat capacity peaks, with no clear sharpening and growth of the peak heights and strong system size dependence as we increase system size. This may indicate that finite-size effects are strong; it is possible that a phase transition only becomes visible on much larger system sizes than we have accessed in our simulations. Further studies are needed to settle this issue. With this caveat in mind, we note that extrapolating the position of the specific heat peak position to the thermodynamic limit gives a putative transition point $T_{c,v}\approx0.390$. As a different probe for the transition temperature, we calculate $\mathcal{M}(T)=\max_{\bk}\sqrt{S(\bk)}$ and normalize it to the range [0,1] in the right panels of Fig. \ref{fig:cvop} for the same phases. In the high-temperature disordered state, the thermal average $\langle \dd_i\rangle=0$ and $\mathcal{M}(T)\rightarrow0$ while in the low-temperature regime the normalized $\mathcal{M}(T)\rightarrow1$. Using $\mathcal{M}$ as an order parameter with the transition temperature defined at $\mathcal{M}(T_c)=0.5$, we find $T_{c,s1}\approx0.609$, $T_{c,s2}\approx0.496$, and $T_{c,s3}\approx0.429$. A similar procedure for the vortex
crystal phase yields $T_{c,v}\approx0.389$, keeping in mind the caveat discussed above. These transition points are in good agreement with the specific heat results.

\section{Fermions on the decorated honeycomb lattice: Coupling to dipoles}
\label{sec:DF}
We next turn to the  electrons moving on the dual decorated honeycomb lattice, which is relevant to various physical realizations 
discussed in the Introduction. The decorated honeycomb lattice generalizes the two-site honeycomb unit cell to $N_s$ sites, as shown in Fig.~\ref{fig:honeycomb}(a) for $N_s=4$. We focus on even $N_s$ since odd $N_s$ leads to flat bands at the $\Gamma$ points at zero energy, which is inconsistent with the known band structures of both conventional and molecular graphene. The unit cell consists of all sites in the Y-shaped wire except for the endpoints on two of the three bonds.
The electron kinetic energy is given by
\begin{equation}
H_e = -\sum_\bk\sum_{\langle a,b\rangle} tc^\dagger_{a}(\bk)c_{b}(\bk)\e^{\i\bk\cdot(\vec{r}_a-\vec{r}_b)}+\mathrm{H.c.},
\label{eqn:He}
\end{equation}
where $a,b\in[1,2,...,3N_s-4]$. Setting the nearest-neighbor dipole distance to unity, the
spacing between neighboring electron sites is $|\vec{r}_{ab}|\equiv\frac{1}{\sqrt{3}}\frac{1}{N_s-1}$.
The fermions couple to the dipoles through their electrostatic potential:
\begin{equation}
H_{e-d} = V\sum_{\bk}\sum_{\langle a,i\rangle}\frac{\dd_i\cdot(\vec{r}_a-\vec{R}_i)}{|\vec{r}_a-\vec{R}_i|^3}c^\dagger_{a}(\bk)c_{a}(\bk),
\label{eqn:Hed}
\end{equation}
where $V$ plays the role of the coupling strength since the dipoles are normalized to unit length in our Monte Carlo (MC) simulations. Here, $\langle a,i\rangle$ means that we only keep the contribution from the three nearest dipoles in order to ensure the full $C_3$ symmetry of the Hamiltonian [see Fig.~\ref{fig:honeycomb}(a)]. The Hamiltonian matrix has dimensions $N_d (3N_s-4)\times N_d (3N_s-4)$ at every momentum $\bk$, where $N_d= L_d \times L_d$ is the number of dipoles we keep in the symmetry broken
unit cell.

\subsection{Non-interacting band structure}
\begin{figure}[t]
	\centering
	\includegraphics[width=\linewidth]{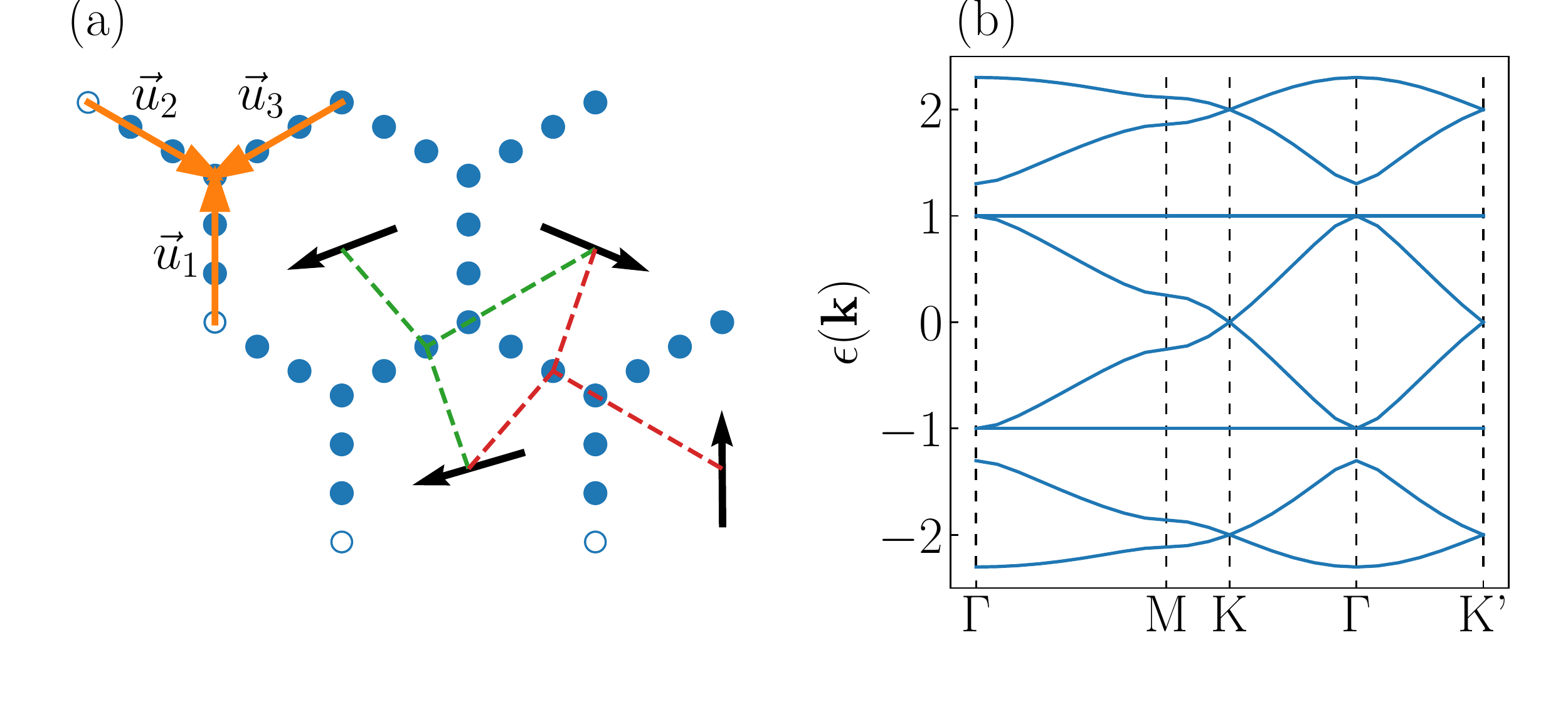}
	\includegraphics[width=\linewidth]{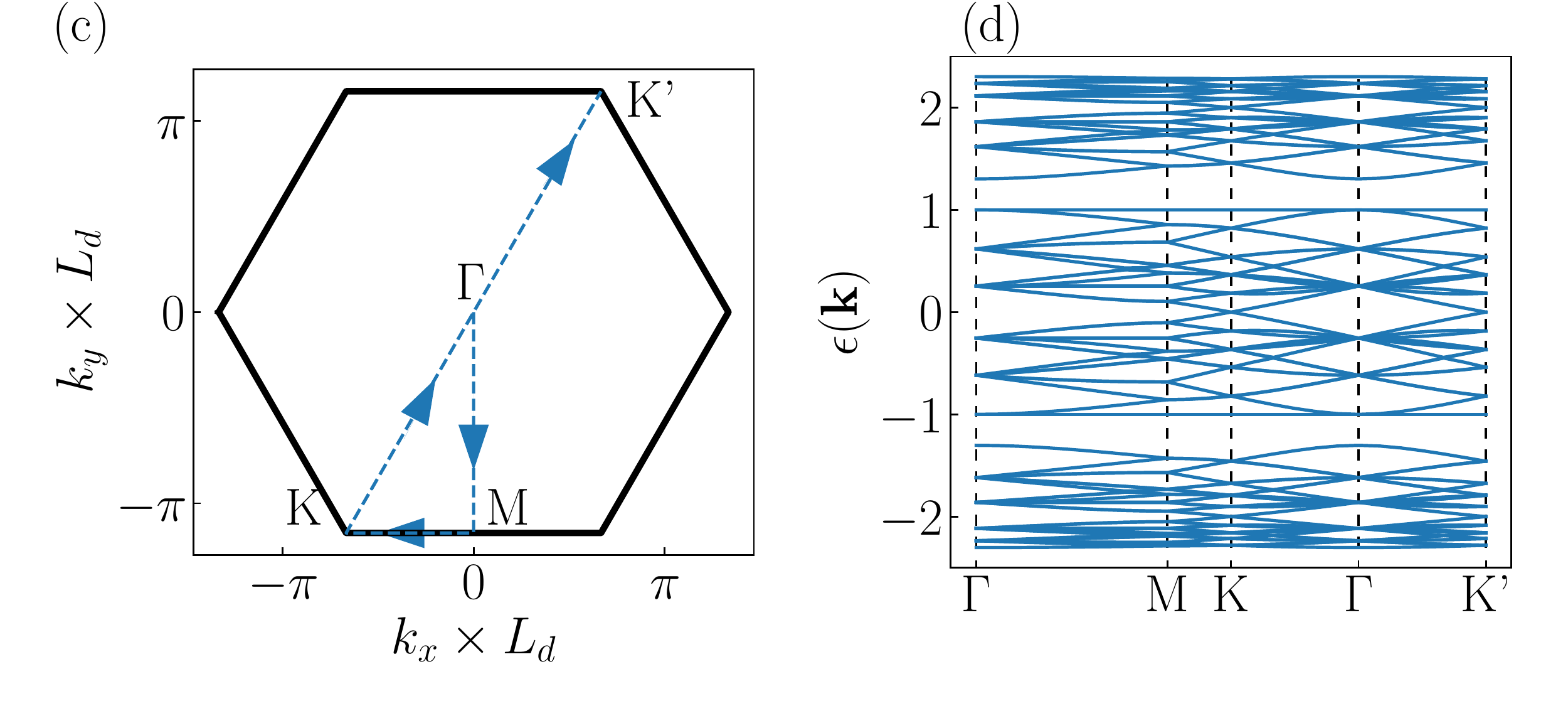}
	\caption{(a) Honeycomb lattice decorated with two additional intermediate sites per bond ($N_s=4$). The black arrows show a generic dipole configuration while the red and green dashed lines denote the three nearest dipole neighbors of two different fermionic sites. The orange arrows show the honeycomb lattice vectors $\vec{u}_{1,2,3}$. (b) Band structure along high symmetry lines of the first Brillouin zone with no dipoles ($V=0$). (c) Reduced Brillouin zone (RBZ) for the supercell consisting of $L_d\times L_d$ dipoles and $N_s$ intermediate sites on the honeycomb lattice. (d) Band structure along high-symmetry lines with $L_d=4$; the additional bands originate from the folding onto the RBZ.}
	\label{fig:honeycomb}
\end{figure}

Without the dipoles, the presence of intermediate sites on the honeycomb lattice leads to new Dirac crossings and 
$(N_s-2)$ flat bands in the electronic spectrum [Fig.~\ref{fig:honeycomb}(b)]. Such non-dispersing bands are known to exist in the Lieb lattice (square lattice equivalent, with $N_s=2$) and they originate in our case from localized wavefunctions on honeycomb rings \cite{honeycombwires}. 
Since odd values of $N_s$ lead to flat bands at zero energy, which is inconsistent with density of states measurements on molecular graphene, 
we focus here and below on even $N_s$ where Dirac band touching appears at zero energy, but 
the flat bands are displaced to positive and negative energies.
When the dipoles are introduced, let us assume that they order into a pattern with the unit cell being enlarged to accommodate
$N_d = L_d\times L_d$ dipoles. Figures .~\ref{fig:honeycomb}(c,d) show the reduced BZ and dispersion,
with the crystal momenta $(k_x,k_y)$ scaled up by 
$L_d$ to make it look like the original BZ. The additional bands in Fig.~\ref{fig:honeycomb}(d) shows how such an enlargement of the unit cell in Real-space leads
to additional bands appearing simply from the folding of bands into the reduced BZ; we have still set $V=0$, so the dipole potential is still zero.

\subsection{Order by disorder for ferrodipolar order}
In the case where all the dipoles tip in-plane and point in the same direction ($\bQ=0$, as with $n_{\max}\rightarrow\infty$), making an angle $\varphi$ with the horizontal axis, there is no preferred direction for the dipoles to point in; all $\varphi$ values have the same energy. We argue that coupling the dipoles to fermions leads to a quantum energy correction which depends on $\varphi$, leading to a six fold clock anisotropy. We illustrate this for the simplest case of $N_s\!=\!2$ (i.e., the 
non decorated honeycomb lattice). In this case, the net potential due to the dipoles vanishes at each site; however on symmetry grounds, we expect
the fermion hopping amplitude $t$ for the nearest-neighbor bonds
to become distinct and functions of $\varphi$. We thus decompose these hoppings as a linear combination of a uniform part with amplitude $t$ and an 
angle-dependent part 
$\delta t(\varphi)$, so it takes the form
\begin{equation}
\begin{pmatrix}
t_1(\varphi)\\
t_2(\varphi)\\
t_3(\varphi)
\end{pmatrix}=t\begin{pmatrix}
1\\
1\\
1
\end{pmatrix}+
\delta t(\varphi)\begin{pmatrix}
1\\
\omega^*\\
\omega
\end{pmatrix}+\delta t^*(\varphi)\begin{pmatrix}
1\\
\omega\\
\omega^*
\end{pmatrix},
\end{equation}
where
$\omega\equiv\e^{\i2\pi/3}$ and $\varphi$ denotes the in-plane angle of the dipoles. Symmetry constraints allow us to set 
${\delta t(\varphi)=|\delta t|\e^{\i 2\varphi}}$ (see Appendix \ref{appendixA}), so that the dipole order acts as a nematic order (even under inversion)
on the electrons. It is convenient to denote $\delta t(\varphi) \equiv \psi$, a nematic order parameter.
The two-band Hamiltonian equation. \eqref{eqn:He} in the presence of ferrodipolar order is
\begin{equation}
H_e(\bk)=-\sum_{l=1}^{3}t_l(\varphi)\e^{\i\bk\cdot\vec{u}_l}c^\dagger_A(\bk)c_B(\bk)+\mathrm{H.c.},
\label{eqn:Hhoneycomb}
\end{equation}
with $A,B$ referring to the sublattice degree of freedom and $\vec{u}_1=\frac{\hat{y}}{\sqrt{3}},\vec{u}_2=\frac{\hat{x}}{2}-\frac{\hat{y}}{2\sqrt{3}}$, and $\vec{u}_3=-\frac{\hat{x}}{2}-\frac{\hat{y}}{2\sqrt{3}}$ as shown in Fig.~\ref{fig:honeycomb}(a). 
Let us define 
\bea
\gamma_1(\bk)&=& \e^{\i k_1}+\e^{\i k_2}+\e^{\i k_3}\\
\gamma_2(\bk)&=& \e^{\i k_1}+\omega^*\e^{\i k_2}+\omega\e^{\i k_3}
\eea
where $k_l\equiv \bk\cdot\vec{u}_l$ (with $l=1,2,3$).
The Hamiltonian can be cast into the matrix form:
\bea
\mathcal{H}_e(\bk)=-th_0(\bk)-\psi h_\psi(\bk)-\psi^* h_{\psi^*}(\bk),
\eea
where the matrices appearing in this equation are given by
\begin{eqnarray}
h_0(\bk)&=&\begin{pmatrix}
0 & \gamma_1(\bk)\\
\gamma_1(-\bk) & 0
\end{pmatrix},\\
h_\psi(\bk)&=&\begin{pmatrix}
0 &  \gamma_2(\bk) \\
\gamma_2(-\bk) & 0
\end{pmatrix},\\
h_{\psi^*}(\bk)&=&\begin{pmatrix}
0 &  \gamma_2^*(-\bk)\\
\gamma_2^*(\bk) & 0
\end{pmatrix}.
\end{eqnarray}
The action after integrating out the fermions is
\begin{equation}
\!\!\! \mathcal{S}\!=\! \Tr\ln\!\left[G_0^{-1}(\bk,\i\omega_n)\!+\!\psi h_\psi(\bk)\!+\! \psi^{*}h_{\psi^*}(\bk)\right]
\end{equation}
where $G^{-1}_0(\bk,\i\omega_n)=\i\omega_n\mathds{1}+th_0(\bk)$ and $\omega_n$ are the fermionic Matsubara frequencies. We expand this
as
${\mathcal{S}=\mathcal{S}_0+\mathcal{S}_{\mathrm{eff}}[\psi,\psi^*]}$ where 
$\mathcal{S}_0=\Tr\ln\left[G_0^{-1}(\bk,\i\omega_n)\right]$ is the bare fermion contribution, and a third-order
perturbative computation of $\mathcal{S}_{\mathrm{eff}}$ yields
\begin{equation}
\mathcal{S}_{\mathrm{eff}}[\psi,\psi^*]=v|\psi|^2+w(\psi^3+\psi^{*3}),
\end{equation}
where both $v$ and $w$ are functions of temperature and density  (see Appendix \ref{appendixA} for details).
We note that this cubic anisotropy for the nematic order corresponds to a $\cos(6\varphi)$
clock anisotropy for the electric dipoles. Thus, depending on the sign of $w$ the ferro-order will pin the dipoles either along or orthogonally to the triangular lattice bonds; our computation yields $w < 0$, which pins the dipoles to point along one of the six nearest-neighbor bond directions of the triangular lattice.

\begin{figure}[!ht]
	\centering
	\includegraphics[width=\linewidth]{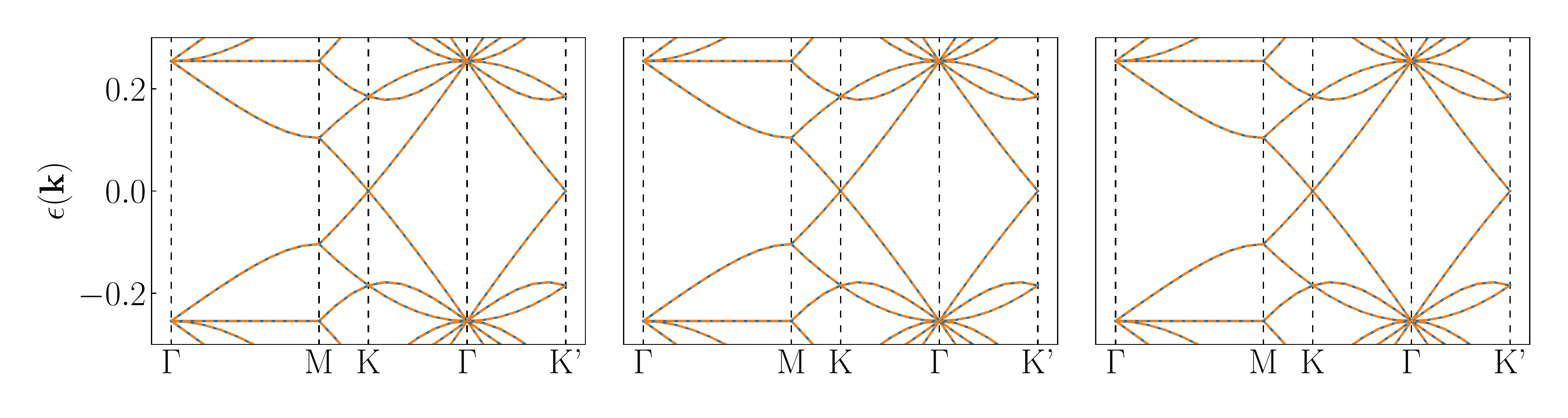}
	\includegraphics[width=\linewidth]{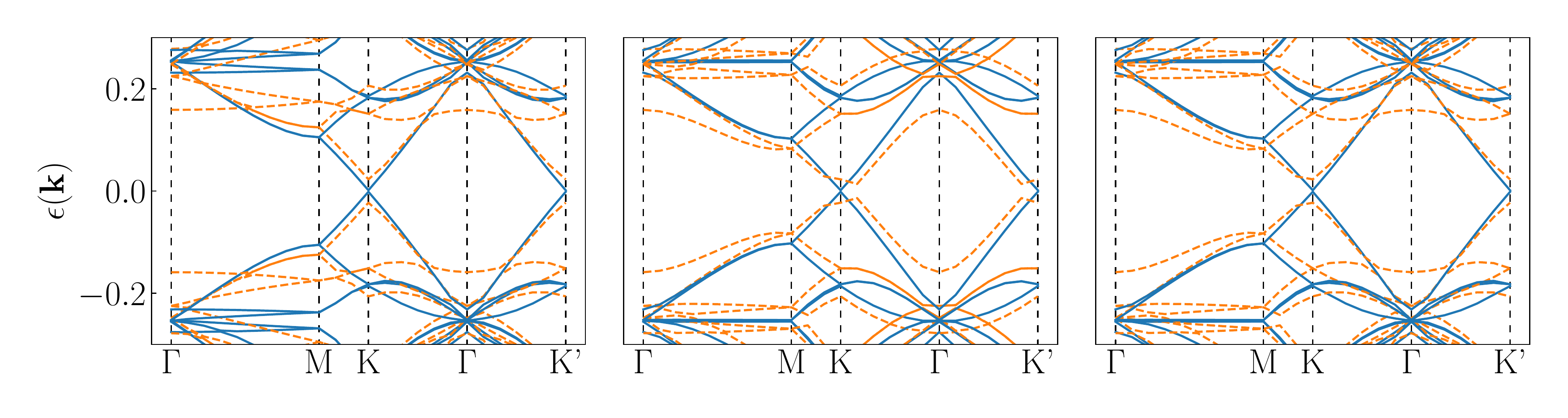}
	\includegraphics[width=\linewidth]{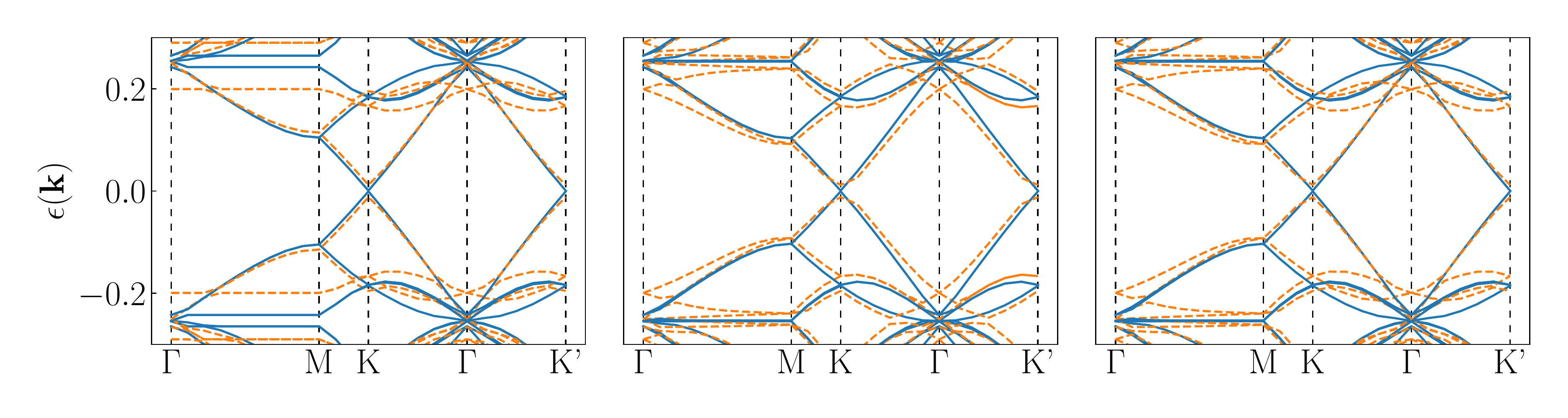}
	\caption[width=\textwidth]{Band structure along high symmetry lines in the RBZ plotted in Fig.~\ref{fig:honeycomb}(c) for the ferro state $|\bQ|=0$ (top row), stripe-1 state $|\bQ|=\mathrm{M}$ (middle row), and stripe-2 state $|\bQ|=\mathrm{M}/2$ (bottom row) using $V=0.1t$ (solid blue curves) and $V=0.4t$ (dashed orange curves) with $L_d=N_s=4$. In each row, we consider dipoles ordered parallel to $\hat{x}$ (left panels), $\hat{x}/2+\sqrt{3}\hat{y}/2$ (middle panels), and
	$\hat{x}/2+\sqrt{3}\hat{y}/2$ (right panels).}
	\label{fig:bs}
\end{figure}

\subsection{Impact of dipole order on the band dispersion}

When the dipoles order,  the underlying symmetries of the triangular lattice are spontaneously broken. This typically leads to
a nonzero gap at the (K,K') points of the RBZ, due to inversion breaking, and causes the flat bands originating from the localized electronic wavefunctions to
become dispersive. The amplitude of such effects depends on the strength of the potential $V$ in Eq. \eqref{eqn:Hed}. 
In addition, the breaking of rotational symmetry means that the
band structure depends on the direction of the dipole orientation. 

In order to illustrate the impact of the dipole potential on the band structure,
we consider the case $N_s = 4$, where different sites on the decorated honeycomb lattice experience different potentials, which allows for the
possibility that the impact of symmetry breaking due to dipole order is more clearly manifest.

Figure~\ref{fig:bs} plots the low-energy part of the band dispersion for $N_s=4$ for the ferrodipolar and two different dipolar stripe 
orders. In these plots, we consider three different orientations for the dipoles, along 
$\hat{x}$ (left panels), ${\hat{x}/2+\sqrt{3}\hat{y}/2}$ (middle panels) and $-\hat{x}/2+\sqrt{3}\hat{y}/2$ (right panels), 
while always choosing the path in the RBZ shown in Fig.~\ref{fig:honeycomb}(c). We plot the dispersions for two different
strengths of the electron dipole coupling, $V=0.1t$ (solid blue curves) and $V=0.4t$ (dashed orange curves).

For the ferrodipolar order (top panels of Fig.~\ref{fig:bs}), we find that the impact of dipoles on the band dispersion is negligible, with a Dirac gap $\sim 10^{-5} t$ from the inversion
breaking. For the case of stripe orders (middle and bottom panels), we find that increasing the dipole coupling from $V=0.1 t$ to $V=0.4 t$ leads to visible
changes in the dispersion, including a more clearly manifest Dirac gap. In addition, we can see signatures of nematicity at higher energies near 
the $\Gamma$ point where $\dd_i\propto\hat{x}\perp\mathrm{M}$ (the left-most panel) leads to a different dispersion from the center and right-most panels.

\subsection{Local density of states}

\begin{figure}[tb]
	\centering
	\includegraphics[width=\linewidth]{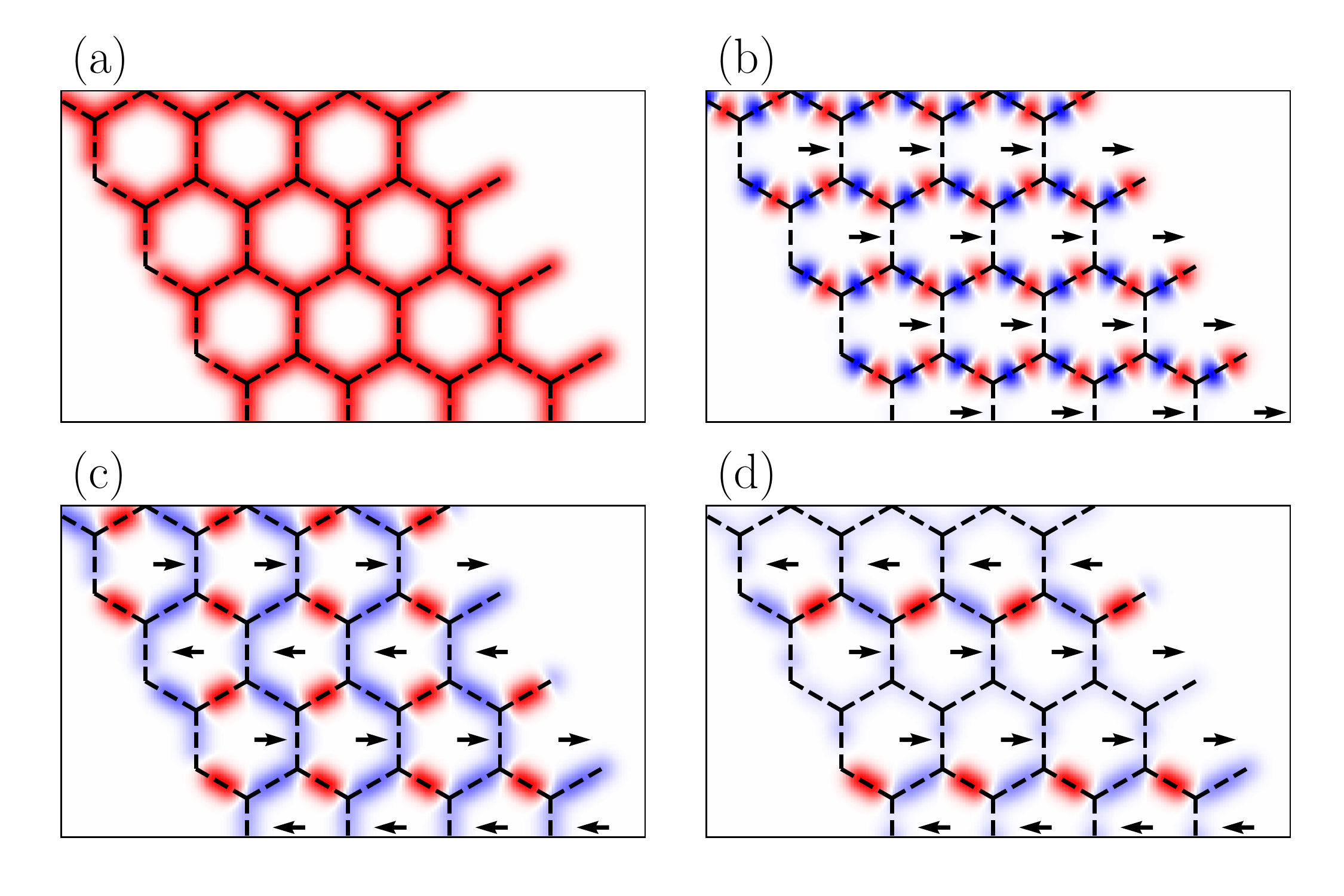}
	\caption{(a) Local density of states near the Dirac points ($E=0.09t$) with red denoting a high LDOS and white denoting a low LDOS in the absence of in-plane dipole moments $\nu_0(\vec{r};E)$, (b) the modulation $\nu_{\mathrm{ferro}}(\vec{r};E)-\nu_0(\vec{r};E)$ with a ferrodipolar order ($|\bQ|=0$), (c) with a stripe-1 order $\nu_{\mathrm{s1}}(\vec{r};E)-\nu_0(\vec{r};E)$ ($|\bQ|=\mathrm{M}$), and (d) with a stripe-2 order $\nu_{\mathrm{s2}}(\vec{r};E)-\nu_0(\vec{r};E)$ ($|\bQ|=\mathrm{M}/2$). Blue denotes negative change in the LDOS while red is a positive change.}
	\label{fig:ldos}
\end{figure}

\begin{figure}[t]
	\centering
	\includegraphics[width=\linewidth]{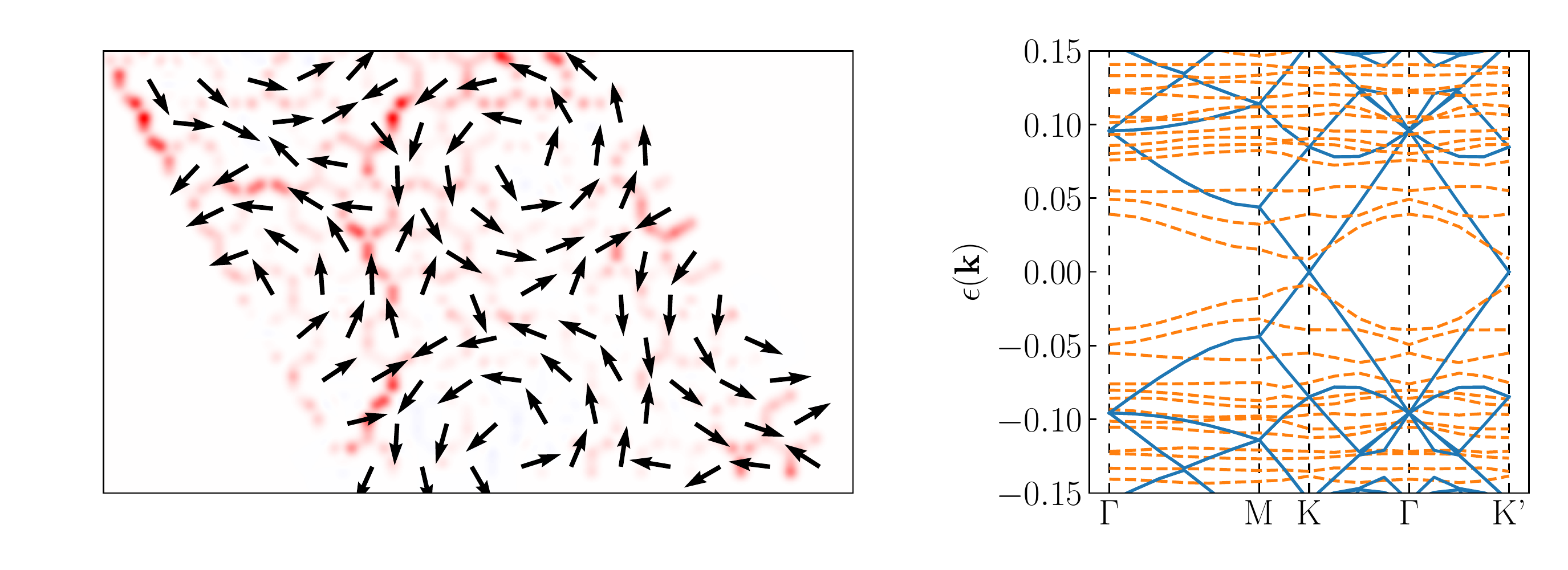}
	\caption{Left panel: Difference between the LDOS in the presence and in the absence of the vortex crystal. The LDOS $\nu_{\mathrm{vortex}}(\vec{r};E)$ for the vortex crystal is calculated for the supercell of $L_d=10$ dipoles at an energy slightly above the Dirac points and with $V=0.8t$. Right panels: Corresponding band structure along high symmetry lines in the RBZ (orange) and with $V=0$ (blue). }
	\label{fig:vortexbs}
\end{figure}

Having examined the effect that the different dipolar orders have on the electronic band structure, we next consider the
local density of states (LDOS) as another probe of the
dipolar ordering. We define the LDOS at a discrete site $a$ on the honeycomb wire and at fixed energy $E$,
\begin{equation}
\nu_a(E)=\sum_\bk \sum_\alpha \mathcal{U}_{a,\alpha}(\bk) \delta[\epsilon_{\alpha}(\bk)-E] \mathcal{U}^\dagger_{\alpha,a}(\bk),
\label{eqn:ldos}
\end{equation}
where $\alpha$ is a band index, $\mathcal{H}(\bk)=\mathcal{H}_e(\bk)+\mathcal{H}_{e-d}(\bk)=\mathcal{U}(\bk)\epsilon(\bk) \mathcal{U}^\dagger(\bk)$ and $\delta[\epsilon_{\alpha}(\bk)-E]$ is the Dirac delta function. In the continuum, we generalize this quantity to any arbitrary position on the substrate by averaging over neighboring sites:
\begin{equation}
\nu(\vec{r};E)=\sum_a \nu_a(E)\e^{-\frac{(\vec{r}-\vec{r}_a)^2}{2c^2}},
\label{eqn:ldosc}
\end{equation}
where the length scale $c$ is chosen to be slightly smaller than the distance between the neighboring decorated sites. In the following,
for $N_s\!=\!4$, the inter-site separation is
$1/(3\sqrt{3})$ (in units of the dipole-dipole distance), and we choose $c=1/(6\sqrt{3})$.
Note that in the limit $c\rightarrow0$, Eq. \eqref{eqn:ldosc} reduces to $\nu(\vec{r}_a;E)=\nu_a(E)$.
We plot $\nu(\vec{r};E)$ at a fixed energy above the Dirac points for the honeycomb lattice for different dipole configurations ($E=0.09t$). In Fig.~\ref{fig:ldos}(a) we show $\nu_0(\vec{r};E)$ with no electron-dipole coupling (equivalent to the case where the dipoles are all pointing along the $z$ axis), and Figs.~\ref{fig:ldos}(b)-\ref{fig:ldos}(d) we plot the modulation in the LDOS for different ordered states. In Fig.~\ref{fig:ldos}(b) we plot $\nu_{\mathrm{ferro}}(\vec{r};E)-\nu_0(\vec{r};E)$ where $\nu_{\mathrm{ferro}}(\vec{r};E)$ is the LDOS for ferro aligned dipoles oriented along the $x$ axis ($\varphi=0$), and we can see that only one mirror plane symmetry parallel to the dipoles is preserved. Although the other symmetries from the $D_{6h}$ point group are broken, a periodicity of the LDOS along the zigzag bonds can be seen, while the modulation on the vertical bond is very weak. Similarly to the ferro order, a mirror plane parallel to the dipoles preserves the lattice symmetry for the stripe-1 order as can be seen in Fig.~\ref{fig:ldos}(c). This mirror plane is generally preserved for stripe-$n$ orders with odd $n$ but is broken for even $n$ such as when second- and third-nearest neighbors are included (which leads to $|\bQ|=\pi/\sqrt{3}$) as shown in Fig.~\ref{fig:ldos}(d). Scanning tunneling spectroscopy (STS) measurements 
could therefore serve as a probe for such ordered dipolar states.

For the vortex crystal (Fig.~\ref{fig:vortexbs}), we find that the LDOS shows a depletion for sites near the core of the vortices, and a corresponding pile-up in the region between vortices. The structure and periodicity of the vortex crystal are reflected in the spatial structure of the LDOS.

Finally, it should be noted that although the amplitude of the LDOS might change depending on the choice of $E$, the symmetry of the signal only depends on the type of dipolar ordering. When $E$ is near the flat bands of the bare band structure, any finite $\bQ$ order disperses the bands and leads to a big drop in the LDOS, making any modulation too small to measure. This is not the case for the $\bQ=0$ order which preserves the flat bands, and the modulation is identical to the one near the Dirac points.

\section{Discussion}

We have shown that dipolar interactions on the triangular lattice with a range cutoff can lead to stripe orders or vortex crystal orders.
Such a reduction from true long-range dipolar couplings to short-range interactions can originate from screening due to the substrate. 
For CO molecules which play the role of the dipoles in molecular graphene, the electric dipole moment is $|\dd|\approx0.02\;\mathrm{e\AA}$ and the lattice constant explored in the original work $a\approx20\;\mathrm{\AA}$ is too large, which leads to a very small coupling constant 
$J_1=|\dd|^2/(4\pi\epsilon_0a^3)\approx1\;\mu \mathrm{eV}$ and transition temperatures $\sim J_1/2\approx 6\;\mathrm{mK}$. However, 
given the $1/a^3$ scaling of the couplings, these small energy scales can be amplified by making the dipolar lattice denser. For instance, a reduction of the inter-dipolar distance by a factor of 10 to $a\approx2\;\mathrm{\AA}$ leads to measurable transition temperatures $\sim 6\;\mathrm{K}$. Using molecules with larger dipole moments would also serve to amplify the scale of the effects we have discussed in this paper.  Scanning tunneling spectroscopy might provide a useful tool to detect such
dipole orders and their influence in molecular graphene and related systems.

\begin{acknowledgments}
This work was funded by NSERC of Canada. N.B. is supported by OGS of Ontario and FRQNT of Quebec. The numerical simulations were performed 
on the Cedar cluster hosted by Westgrid and Compute Canada. We thank Hari Manoharan and Ivar Martin for useful discussions.
\end{acknowledgments}

\bibliography{biblio}
\appendix

\begin{widetext}
\section{Derivation of the order by disorder effective action}
\label{appendixA}

We consider a small modulation of the hopping amplitude on the honeycomb lattice ($N_s=2$) caused by the ferrodipolar order. We write $t_l(\varphi)$ along each of the $l=1,2,3$ honeycomb bonds as a linear combination of a uniform hopping $t$ (i.e., the bare kinetic energy when $V=0$ or $\dd\propto\hat{z}$) and a modulation $\delta t_l(\varphi)$:
\begin{equation}
\begin{pmatrix}
t_1(\varphi)\\
t_2(\varphi)\\
t_3(\varphi)
\end{pmatrix}=t\begin{pmatrix}
1\\
1\\
1
\end{pmatrix}+\begin{pmatrix}
\delta t_1(\varphi)\\
\delta t_2(\varphi)\\
\delta t_3(\varphi)
\end{pmatrix},
\end{equation}
Since the modulated piece is orthogonal to the $(1,1,1)^\top$ subspace, for real hoppings this can be re-written as:
\begin{equation}
\begin{pmatrix}
\delta t_1(\varphi)\\
\delta t_2(\varphi)\\
\delta t_3(\varphi)
\end{pmatrix}=\delta t(\varphi)\begin{pmatrix}
1\\
\omega^*\\
\omega
\end{pmatrix}+\delta t^*(\varphi)\begin{pmatrix}
1\\
\omega\\
\omega^*
\end{pmatrix},
\label{eqn:hopping}
\end{equation}
with $\omega\equiv\e^{\i2\pi/3}$ and $\delta t(\varphi)=|\delta t|\e^{\i f(\varphi)}$. Our goal is to find $f(\varphi)$ from symmetry constraints. For example, when $\varphi=0$ we expect $t_2(\varphi=0)=t_3(\varphi=0)$ [see Fig.~\ref{fig:honeycomb}(a)] and when $\varphi=\pi/3$ we expect $t_1(\varphi=\pi/3)=t_3(\varphi=\pi/3)$. From these as well as from similar constraints when $\varphi=2\pi/3$ and $\varphi=\pi/2$ we can deduce:

\begin{eqnarray}
\delta t_2\left(0\right) &=& \delta t_3\left(0\right) \implies \cos\left(f\left(0\right)-\frac{2\pi}{3}\right)=\cos\left(f\left(0\right)+\frac{2\pi}{3}\right)\iff f\left(0\right)=n\pi,\\
\delta t_1\left(\frac{\pi}{3}\right) &=& \delta t_3\left(\frac{\pi}{3}\right) \implies \cos\left[f\left(\frac{\pi}{3}\right)\right]=\cos\left[f\left(\frac{\pi}{3}\right)+\frac{2\pi}{3}\right]\iff f\left(\frac{\pi}{3}\right)=n\pi-\frac{\pi}{3},\\
\delta t_1\left(\frac{2\pi}{3}\right) &=& \delta t_2\left(\frac{2\pi}{3}\right) \implies \cos\left[f\left(\frac{2\pi}{3}\right)\right]=\cos\left[f\left(\frac{2\pi}{3}\right)-\frac{2\pi}{3}\right] \iff f\left(\frac{2\pi}{3}\right)=n\pi-\frac{2\pi}{3},\\
\delta t_2\left(\frac{\pi}{2}\right) &=& \delta t_3\left(\frac{\pi}{2}\right) \implies \cos\left[f\left(\frac{\pi}{2}\right)-\frac{2\pi}{3}\right]=\cos\left[f\left(\frac{\pi}{2}\right)+\frac{2\pi}{3}\right] \iff f\left(\frac{\pi}{2}\right)=n\pi, \\
\mathrm{etc.,}\nonumber
\end{eqnarray}
where $n\in\mathbb{Z}$. The simplest choice of $f(\varphi)$ which satisfies all the constraints is $f(\varphi)=2\varphi$. We construct the nematic order parameter $\psi=\delta t(\varphi)=|\delta t|\;\e^{2\i\varphi}$ and rewrite Eq. \eqref{eqn:hopping} as:
\begin{equation}
\begin{pmatrix}
\delta t_1(\varphi)\\
\delta t_2(\varphi)\\
\delta t_3(\varphi)
\end{pmatrix}=\psi\begin{pmatrix}
1\\
\omega^*\\
\omega
\end{pmatrix}+\psi^*\begin{pmatrix}
1\\
\omega\\
\omega^*
\end{pmatrix},
\label{eqn:psi}
\end{equation}
After substituting this in the Hamiltonian \eqref{eqn:Hhoneycomb}, we write the partition function as a path integral over fermion fields $(c,\bar{c})$ and the nematic $(\psi,\psi^*)$ and integrate out the fermions:
\begin{eqnarray*}
	Z&=&\int\mathcal{D}(\psi,\psi^*)\int\mathcal{D}(c,\bar{c})\;\exp\left[-\sum_{\bk,\omega_n}\bar{c}(\bk)[G_0^{-1}(\bk,\i\omega_n)+\psi h_\psi(\bk)+\psi^*h_{\psi^*}(\bk)]c(\bk)\right]\\
	&=&\int\mathcal{D}(\psi,\psi^*)\e^{-\Tr\ln\left[G_0^{-1}(\bk,\i\omega_n)+\psi h_\psi(\bk)+\psi^*h_{\psi^*}(\bk)\right]}.
\end{eqnarray*}
We evaluate the action $\mathcal{S}[\psi,\psi^*]$ perturbatively in powers of $\psi$ and $\psi^*$:
\begin{eqnarray}
\mathcal{S}[\psi,\psi^*]&=&\Tr\ln\left[G_0^{-1}(\bk,\i\omega_n)+\psi h_\psi(\bk)+\psi^*h_{\psi^*}(\bk)\right]\nonumber\\
&=&\Tr\ln\left[G_0^{-1}(\bk,\i\omega_n)\left(\mathds{1}+\psi G_0(\bk,\i\omega_n) h_\psi(\bk)+\psi^*G_0(\bk,\i\omega_n)h_{\psi^*}(\bk)\right)\right]\nonumber\\
&=&\mathcal{S}_0+\Tr\sum_m \frac{(-1)^m}{m} \left(\psi G_0(\bk,\i\omega_n) h_\psi(\bk)+\psi^*G_0(\bk,\i\omega_n)h_{\psi^*}(\bk)\right)^m.
\end{eqnarray}
We will consider the $m=1,2,3$ contributions to generate up to $\psi^3$ and $\psi^{*3}$ terms:
\begin{eqnarray}
\mathcal{S}_{\mathrm{eff}}[\psi,\psi^*]&\simeq&-\psi\Tr\left(G_0h_\psi\right)-\psi^*\Tr\left(G_0h_{\psi^*}\right)\nonumber\\
&+&\frac{1}{2}\left[\psi^2 \Tr\left(\left(G_0h_\psi\right)^2\right) + 2|\psi|^2 \Tr\left(G_0h_\psi G_0h_{\psi^*}\right)  + \psi^{*2} \Tr\left(\left(G_0h_{\psi^*}\right)^2\right)\right]\nonumber\\
&-&\frac{1}{3}\left[\psi^3 \Tr\left(\left(G_0h_\psi\right)^3\right) + 3|\psi|^2\psi \Tr\left(\left(G_0h_\psi\right)^2G_0h_{\psi^*}\right)  +  3|\psi|^2\psi^* \Tr\left(\left(G_0h_{\psi^*}\right)^2G_0h_\psi\right) + \psi^{*3} \Tr\left(\left(G_0h_{\psi^*}\right)^3\right)\right].\nonumber
\end{eqnarray}
Here we dropped the dependence on momenta $\bk$ and fermionic Matsubara frequencies $\omega_n=(2n+1)\pi/\beta$ for conciseness. The first order correction is straightforward to calculate:
\begin{eqnarray}
\Tr\left(G_0h_\psi\right)&=&\Tr\left(\left(\i\omega_n+th_0\right)^{-1}h_\psi\right)=\Tr\left(U\frac{1}{\i\omega_n-\xi}U^\dagger h_\psi\right)\nonumber\\
&=&\sum_{\bk,\i\omega_n} \sum_{\mu,\nu,\alpha} U_\bk^{\mu,\nu}\frac{1}{\i\omega_n-\xi_\bk^\nu}U_\bk^{\dagger\nu,\alpha}h_{\psi,\bk}^{\alpha,\mu}\nonumber\\
&=&\sum_\bk \beta \tr\left[U_\bk n_F(\xi_\bk)U^\dagger_\bk h_{\psi,\bk}\right],
\end{eqnarray}
and similarly:
\begin{eqnarray}
\Tr\left(G_0h_{\psi^*}\right)&=&\sum_\bk \beta \tr\left[U_\bk n_F(\xi_\bk)U^\dagger_\bk h_{\psi^*,\bk}\right],
\end{eqnarray}
where $\tr$ is the conventional matrix trace (as opposed to $\Tr$ the trace over momenta, Matsubara frequencies, and sublattice degrees of freedom). $\beta=1/T$ is the inverse temperature and $n_F(\xi)=(1+\e^{\beta \xi})^{-1}$ is the Fermi-Dirac distribution. $U_\bk$ is the matrix which diagonalizes the free electron Hamiltonian $\mathcal{H}_{e,0}=-th_0=U\epsilon U^\dagger$ and $\xi=\epsilon-\mu$ are the energy eigenvalues with respect to the chemical potential $\mu$ for a fixed electron density. The second-order correction can be evaluated in a similar fashion:
\begin{eqnarray}
\Tr\left(\left(G_0h_\psi\right)^2\right)&=&\Tr\left(\left(\i\omega_n+th_0\right)^{-1}h_\psi\left(\i\omega_n+th_0\right)^{-1}h_\psi\right)=\Tr\left(U\frac{1}{\i\omega_n-\xi}U^\dagger h_\psi U\frac{1}{\i\omega_n-\xi}U^\dagger h_\psi\right)\nonumber\\
&=&\sum_{\bk,\i\omega_n} \sum_{\mu,\nu,\alpha} \sum_{\gamma,\rho,\sigma} U_\bk^{\mu,\nu}\frac{1}{\i\omega_n-\xi_\bk^\nu}U_\bk^{\dagger\nu,\alpha}h_{\psi,\bk}^{\alpha,\gamma} U_\bk^{\gamma,\rho}\frac{1}{\i\omega_n-\xi_\bk^\rho}U_\bk^{\dagger\rho,\sigma}h_{\psi,\bk}^{\sigma,\mu}\nonumber
\end{eqnarray}
To evaluate the Matsubara frequency summations, we need to distinguish two cases in this two-band problem:
\begin{eqnarray}
\mathcal{A}_\bk^{\nu,\rho}\equiv\mathcal{A}(\xi_\bk^\nu,\xi_\bk^\rho)=\sum_{\i\omega_n} \frac{1}{\i\omega_n-\xi_\bk^\nu}\frac{1}{\i\omega_n-\xi_\bk^\rho}=\begin{cases}
\beta \dfrac{n_F(\xi_\bk^\nu)-n_F(\xi_\bk^\rho)}{\xi_\bk^\nu-\xi_\bk^\rho}\hspace{2.7cm} \mathrm{if\;} \xi_\bk^\nu\neq\xi_\bk^\rho\\
\beta \dfrac{\partial n_F(\xi_\bk^\nu)}{\partial \xi_\bk^\nu} = -\beta^2 \dfrac{\e^{\beta\xi_\bk^\nu}}{(1+\e^{\beta\xi_\bk^\nu})^2} \hspace{1.1cm} \mathrm{if\;} \xi_\bk^\nu=\xi_\bk^\rho,
\end{cases}
\end{eqnarray}
such that:
\begin{eqnarray}
\Tr\left(\left(G_0h_\psi\right)^2\right)&=&\sum_{\bk} \sum_{\mu,\nu,\alpha} \sum_{\gamma,\rho,\sigma} U_\bk^{\mu,\nu}U_\bk^{\dagger\nu,\alpha}h_{\psi,\bk}^{\alpha,\gamma} U_\bk^{\gamma,\rho}U_\bk^{\dagger\rho,\sigma}h_{\psi,\bk}^{\sigma,\mu} \mathcal{A}_\bk^{\nu,\rho},\nonumber\\
\Tr\left(G_0h_\psi G_0h_{\psi^*}\right)&=&\sum_{\bk} \sum_{\mu,\nu,\alpha} \sum_{\gamma,\rho,\sigma} U_\bk^{\mu,\nu}U_\bk^{\dagger\nu,\alpha}h_{\psi,\bk}^{\alpha,\gamma} U_\bk^{\gamma,\rho}U_\bk^{\dagger\rho,\sigma}h_{\psi^*,\bk}^{\sigma,\mu} \mathcal{A}_\bk^{\nu,\rho},\nonumber\\
\Tr\left(\left(G_0h_{\psi^*}\right)^2\right)&=&\sum_{\bk} \sum_{\mu,\nu,\alpha} \sum_{\gamma,\rho,\sigma} U_\bk^{\mu,\nu}U_\bk^{\dagger\nu,\alpha}h_{\psi^*,\bk}^{\alpha,\gamma} U_\bk^{\gamma,\rho}U_\bk^{\dagger\rho,\sigma}h_{\psi^*,\bk}^{\sigma,\mu} \mathcal{A}_\bk^{\nu,\rho}\nonumber.
\end{eqnarray}
Lastly, we compute the third order contributions:
\begin{eqnarray}
\Tr\left(\left(G_0h_\psi\right)^3\right)&=&\Tr\left(\left(\i\omega_n+th_0\right)^{-1}h_\psi\left(\i\omega_n+th_0\right)^{-1}h_\psi\left(\i\omega_n+th_0\right)^{-1}h_\psi\right)\nonumber\\
&=&\Tr\left(U\frac{1}{\i\omega_n-\xi}U^\dagger h_\psi U\frac{1}{\i\omega_n-\xi}U^\dagger h_\psi U\frac{1}{\i\omega_n-\xi}U^\dagger h_\psi\right)\nonumber\\
&=&\sum_{\bk,\i\omega_n} \sum_{\mu,\nu,\alpha} \sum_{\gamma,\rho,\sigma} \sum_{\lambda,\eta,\chi} U_\bk^{\mu,\nu}\frac{1}{\i\omega_n-\xi_\bk^\nu}U_\bk^{\dagger\nu,\alpha}h_{\psi,\bk}^{\alpha,\gamma} U_\bk^{\gamma,\rho}\frac{1}{\i\omega_n-\xi_\bk^\rho}U_\bk^{\dagger\rho,\sigma}h_{\psi,\bk}^{\sigma,\lambda} U_\bk^{\lambda,\eta}\frac{1}{\i\omega_n-\xi_\bk^\eta}U_\bk^{\dagger\eta,\chi}h_{\psi,\bk}^{\chi,\mu}\nonumber.
\end{eqnarray}
Since we have a $2\times2$ Hamiltonian and three energy eigenvalues $\xi_\bk^\nu,\xi_\bk^\rho,\xi_\bk^\eta$, at least two energies must be equal:
\begin{eqnarray}
\mathcal{B}_\bk^{\nu,\rho,\eta}\equiv\mathcal{B}(\xi_\bk^\nu,\xi_\bk^\rho,\xi_\bk^\eta)=\sum_{\i\omega_n} \frac{1}{\i\omega_n-\xi_\bk^\nu}\frac{1}{\i\omega_n-\xi_\bk^\rho}\frac{1}{\i\omega_n-\xi_\bk^\eta}=\begin{cases}
\sum_{\i\omega_n} \dfrac{1}{\left(\i\omega_n-\xi_\bk^\nu\right)^3}\hspace{1.95cm} \mathrm{if\;} \xi_\bk^\nu=\xi_\bk^\rho=\xi_\bk^\eta,\\
\sum_{\i\omega_n} \dfrac{1}{\left(\i\omega_n-\xi_\bk^\nu\right)^2}\dfrac{1}{\i\omega_n-\xi_\bk^\eta} \hspace{0.6cm}\mathrm{if\;} \xi_\bk^\nu=\xi_\bk^\rho\neq\xi_\bk^\eta,\\
\sum_{\i\omega_n} \dfrac{1}{\left(\i\omega_n-\xi_\bk^\nu\right)^2}\dfrac{1}{\i\omega_n-\xi_\bk^\rho} \hspace{0.6cm}\mathrm{if\;} \xi_\bk^\nu=\xi_\bk^\eta\neq\xi_\bk^\rho,\\
\sum_{\i\omega_n} \dfrac{1}{\left(\i\omega_n-\xi_\bk^\rho\right)^2}\dfrac{1}{\i\omega_n-\xi_\bk^\nu} \hspace{0.6cm}\mathrm{if\;} \xi_\bk^\rho=\xi_\bk^\eta\neq\xi_\bk^\nu.
\end{cases}
\end{eqnarray}
In general, we have
\begin{eqnarray}
\sum_{\i\omega_n}\frac{1}{(\i\omega_n-\varepsilon_1)^3}=\frac{\beta}{2}\frac{\partial^2n_F(\varepsilon_1)}{\partial\varepsilon^2}=\frac{\beta^3\e^{2\beta\varepsilon_1}}{(1+\e^{\beta\varepsilon_1})^3}-\frac{1}{2}\frac{\beta^3\e^{\beta\varepsilon_1}}{(1+\e^{\beta\varepsilon_1})^2},
\end{eqnarray}
and for $\varepsilon_1\neq\varepsilon_2$,
\begin{eqnarray}
\sum_{\i\omega_n}\frac{1}{(\i\omega_n-\varepsilon_1)^2}\frac{1}{\i\omega_n-\varepsilon_2}&=&\sum_{\i\omega_n}\frac{1}{\varepsilon_1-\varepsilon_2}\frac{1}{(\i\omega_n-\varepsilon_1)^2}-\frac{1}{(\varepsilon_1-\varepsilon_2)^2}\frac{1}{\i\omega_n-\varepsilon_1}+\frac{1}{(\varepsilon_1-\varepsilon_2)^2}\frac{1}{\i\omega_n-\varepsilon_2}\nonumber\\
&=&\frac{\beta}{\varepsilon_1-\varepsilon_2}\frac{\partial n_F(\varepsilon_1)}{\partial\varepsilon}+\frac{\beta}{\varepsilon_1-\varepsilon_2}\frac{n_F(\varepsilon_2)-n_F(\varepsilon_1)}{\varepsilon_1-\varepsilon_2}\nonumber\\
&=&-\frac{\beta^2\e^{\beta\varepsilon_1}}{(\varepsilon_1-\varepsilon_2)(1+\e^{\beta\varepsilon_1})^2}+\frac{\beta}{(\varepsilon_1-\varepsilon_2)^2}\left[\frac{1}{1+\e^{\beta\varepsilon_2}}-\frac{1}{1+\e^{\beta\varepsilon_1}}\right],
\end{eqnarray}
which leads to the closed form for $\mathcal{B}$:
\begin{eqnarray}
\mathcal{B}_\bk^{\nu,\rho,\eta}=\begin{cases}
\dfrac{\beta^3\e^{2\beta\xi_\bk^\nu}}{(1+\e^{\beta\xi_\bk^\nu})^3}-\dfrac{1}{2}\dfrac{\beta^3\e^{\beta\xi_\bk^\nu}}{(1+\e^{\beta\xi_\bk^\nu})^2}\hspace{6.4cm} \mathrm{if\;} \xi_\bk^\nu=\xi_\bk^\rho=\xi_\bk^\eta\\
-\dfrac{\beta^2\e^{\beta\xi_\bk^\nu}}{(\xi_\bk^\nu-\xi_\bk^\eta)(1+\e^{\beta\xi_\bk^\nu})^2}+\dfrac{\beta}{(\xi_\bk^\nu-\xi_\bk^\eta)^2}\left[\dfrac{1}{1+\e^{\beta\xi_\bk^\eta}}-\dfrac{1}{1+\e^{\beta\xi_\bk^\nu}}\right] \hspace{1.6cm}\mathrm{if\;} \xi_\bk^\nu=\xi_\bk^\rho\neq\xi_\bk^\eta\\
-\dfrac{\beta^2\e^{\beta\xi_\bk^\nu}}{(\xi_\bk^\nu-\xi_\bk^\rho)(1+\e^{\beta\xi_\bk^\nu})^2}+\dfrac{\beta}{(\xi_\bk^\nu-\xi_\bk^\rho)^2}\left[\dfrac{1}{1+\e^{\beta\xi_\bk^\rho}}-\dfrac{1}{1+\e^{\beta\xi_\bk^\nu}}\right] \hspace{1.6cm}\mathrm{if\;} \xi_\bk^\nu=\xi_\bk^\eta\neq\xi_\bk^\rho\\
-\dfrac{\beta^2\e^{\beta\xi_\bk^\rho}}{(\xi_\bk^\rho-\xi_\bk^\nu)(1+\e^{\beta\xi_\bk^\rho})^2}+\dfrac{\beta}{(\xi_\bk^\rho-\xi_\bk^\nu)^2}\left[\dfrac{1}{1+\e^{\beta\xi_\bk^\nu}}-\dfrac{1}{1+\e^{\beta\xi_\bk^\rho}}\right] \hspace{1.6cm}\mathrm{if\;} \xi_\bk^\rho=\xi_\bk^\eta\neq\xi_\bk^\nu.
\end{cases}
\end{eqnarray}
We can finally write:
\begin{eqnarray}
\Tr\left(\left(G_0h_\psi\right)^3\right)&=&\sum_{\bk} \sum_{\mu,\nu,\alpha} \sum_{\gamma,\rho,\sigma} \sum_{\lambda,\eta,\chi} U_\bk^{\mu,\nu}U_\bk^{\dagger\nu,\alpha}h_{\psi,\bk}^{\alpha,\gamma} U_\bk^{\gamma,\rho}U_\bk^{\dagger\rho,\sigma}h_{\psi,\bk}^{\sigma,\lambda} U_\bk^{\lambda,\eta}U_\bk^{\dagger\eta,\chi}h_{\psi,\bk}^{\chi,\mu}\mathcal{B}^{\nu,\rho,\eta},\\
\Tr\left(\left(G_0h_\psi\right)^2G_0h_{\psi^*}\right)&=&\sum_{\bk} \sum_{\mu,\nu,\alpha} \sum_{\gamma,\rho,\sigma} \sum_{\lambda,\eta,\chi} U_\bk^{\mu,\nu}U_\bk^{\dagger\nu,\alpha}h_{\psi,\bk}^{\alpha,\gamma} U_\bk^{\gamma,\rho}U_\bk^{\dagger\rho,\sigma}h_{\psi,\bk}^{\sigma,\lambda} U_\bk^{\lambda,\eta}U_\bk^{\dagger\eta,\chi}h_{\psi^*,\bk}^{\chi,\mu}\mathcal{B}^{\nu,\rho,\eta},\\
\Tr\left(\left(G_0h_{\psi^*}\right)^2G_0h_\psi\right)&=&\sum_{\bk} \sum_{\mu,\nu,\alpha} \sum_{\gamma,\rho,\sigma} \sum_{\lambda,\eta,\chi} U_\bk^{\mu,\nu}U_\bk^{\dagger\nu,\alpha}h_{\psi^*,\bk}^{\alpha,\gamma} U_\bk^{\gamma,\rho}U_\bk^{\dagger\rho,\sigma}h_{\psi^*,\bk}^{\sigma,\lambda} U_\bk^{\lambda,\eta}U_\bk^{\dagger\eta,\chi}h_{\psi,\bk}^{\chi,\mu}\mathcal{B}^{\nu,\rho,\eta},\\
\Tr\left(\left(G_0h_{\psi^*}\right)^3\right)&=&\sum_{\bk} \sum_{\mu,\nu,\alpha} \sum_{\gamma,\rho,\sigma} \sum_{\lambda,\eta,\chi} U_\bk^{\mu,\nu}U_\bk^{\dagger\nu,\alpha}h_{\psi^*,\bk}^{\alpha,\gamma} U_\bk^{\gamma,\rho}U_\bk^{\dagger\rho,\sigma}h_{\psi^*,\bk}^{\sigma,\lambda} U_\bk^{\lambda,\eta}U_\bk^{\dagger\eta,\chi}h_{\psi^*,\bk}^{\chi,\mu}\mathcal{B}^{\nu,\rho,\eta}.
\end{eqnarray}
For a fixed electron density $\bar{n}$, we self-consistently calculate the chemical potential and compute the different contributions as a function of temperature $T=1/\beta$. We find that for all densities and temperatures,
\begin{eqnarray}
\begin{cases}
\Tr\left(G_0h_{\psi}\right)=0\\
\Tr\left(G_0h_{\psi^*}\right)=0\\
\Tr\left(\left(G_0h_\psi\right)^2\right)=0\\
\Tr\left(\left(G_0h_{\psi^*}\right)^2\right)=0\\
\Tr\left(\left(G_0h_\psi\right)^2G_0h_{\psi^*}\right)=0\\
\Tr\left(\left(G_0h_{\psi^*}\right)^2G_0h_\psi\right)=0,
\end{cases}
\end{eqnarray}
such that the effective action reduces to the compact form:
\begin{equation}
\mathcal{S}_{\mathrm{eff}}[\psi,\psi^*]=v|\psi|^2+w(\psi^3+\psi^{*3}),
\end{equation}
where
\begin{eqnarray}
\begin{cases}
v=\Tr\left(G_0h_\psi G_0h_{\psi^*}\right)\\
w=-\frac{1}{3}\Tr\left(\left(G_0h_\psi\right)^3\right)=-\frac{1}{3}\Tr\left(\left(G_0h_{\psi^*}\right)^3\right).
\end{cases}
\end{eqnarray}
Numerical calculations show $v<0, w<0$. Since $\psi^3+\psi^{*3}\propto \cos(6\varphi)$, this indicates that the action is minimized when $\varphi=n\pi/3$ with $n=0,1,...,5$ and, as such, that the dipole moments will be pinned along the triangular lattice nearest-neighbor directions.
\end{widetext}

\end{document}